\documentclass[11pt,a4paper,singlecolumn]{article}
\usepackage[utf8]{inputenc}
\usepackage{makecell}
\usepackage{color,xcolor, url, float,  verbatim, changepage}
\usepackage{graphicx,setspace}
\usepackage[linesnumbered,vlined,ruled]{algorithm2e}
\usepackage{array, multirow, tabu} 
\usepackage{mathtools,amsmath,amssymb} 
\usepackage{hyperref}
\usepackage{multirow}
\usepackage{authblk,stackengine}
\usepackage{colortbl}
\usepackage{enumitem}  
\usepackage{tikz}
\usetikzlibrary{arrows.meta, positioning, fit}

\usepackage{lmodern}
\usepackage[T1]{fontenc}
\usepackage{textcomp}
\usepackage{booktabs}
\usepackage{caption}
\usepackage{natbib} 
\usepackage{authblk,stackengine}
\usepackage{colortbl}
\usepackage{titlesec}
\usepackage{amsthm}

\definecolor{denim}{rgb}{0.08, 0.38, 0.74}
\definecolor{electricindigo}{rgb}{0.44, 0.0, 1.0}
\definecolor{electricultramarine}{rgb}{0.25, 0.0, 1.0}

\usepackage{xcolor} 
\usepackage{soul} 

\usepackage{pifont}  
\newcommand{\cmark}{\ding{51}}
\newcommand{\xmark}{\ding{55}}

\usepackage{tabularx}

\usepackage{array}
\newcolumntype{C}[1]{>{\centering\arraybackslash}m{#1}}
\newcolumntype{y}[1]{>{\raggedright\arraybackslash}m{#1}}
\newcolumntype{Y}{>{\raggedright\arraybackslash}X}
\usepackage[margin = 25mm]{geometry}
\renewcommand{\arraystretch}{1.4}  

\titlespacing*{\subsection}{0pt}{1.5ex}{0.5ex}
\titlespacing*{\subsubsection}{0pt}{1.5ex}{0.5ex}
\providecommand{\keywords}[1]
{
  \small	
  \textbf{\textit{Keywords---}} #1
}

\title{Optimization Under Uncertainty for Energy Infrastructure Planning: A Synthesis of Methods, Tools, and Open Challenges}
\author{Rahman Khorramfar $^{1,2}$\footnote{Corresponding author:khorram@mit.edu}$^*$, Aron Brenner $^{2,3}$, Lara Booth $^{2,4}$, Ana Rivera $^{2,4}$, Ruaridh Macdonald $^1$, Priya Donti $^{2,4}$, Saurabh Amin $^{2,3}$    \\
        \small $^{1}$ MIT Energy Initiative, MIT \\
        \small $^{2}$ Laboratory for Information and Decision Systems (LIDS), MIT \\
        \small $^3$ Civil and Environmental Engineering (CEE), MIT \\
        \small $^4$ Electrical Engineering and Computer Science (EECS), MIT}       
\date{}

\begin{document}
\maketitle
\begin{abstract}
Energy infrastructure planning under uncertainty has become increasingly complex as electrification, interdependence between energy carriers, decarbonization, and extreme weather events reshape long-term investment decisions. This paper surveys recent advances at the intersection of generation and transmission expansion, and optimization under uncertainty, with a focus on stochastic programming, robust optimization, and distributionally robust optimization. We then categorize modeling needs along the axes of modeling fidelity, uncertainty characterization, and solution methods to identify dominant modeling features and trace research gaps. We further examine emerging directions at the interface of optimization and machine learning, including {forecasting, scenario generation, and learned surrogates} and discuss how these tools can be embedded within infrastructure planning models. %
\end{abstract}


\section{Introduction}
The global energy landscape is undergoing a profound transformation aimed at drastically reducing greenhouse gas emissions to mitigate the worst impacts of climate change. This transformation, often referred to as \textit{energy transition}, entails a gradual shift from fossil fuels to low- or zero-carbon energy systems while meeting certain sustainability targets such as resiliency and energy affordability. The transition has accelerated in recent years due to the declining cost of renewable technologies, advances in energy efficiency and storage, growing social awareness,  and digitalization of energy management and distribution \citep{IRENA2019}. Nevertheless, the current pace of change remains insufficient to meet global climate objectives, including those outlined in the Paris Agreement \citep{ParisAgreement2015}. Achieving these objectives requires an overhaul of energy infrastructure through long-term planning efforts that account for regional disparities, sector-specific challenges, operational feasibility, and the often competing interests of different key stakeholders.

Planning for energy infrastructure is facilitated by capacity expansion models (CEMs) that determine the optimal timing, location, and scale of major infrastructure assets, including electric power generators, electric transmission lines, gas pipelines, and grid-scale storage facilities. These models are critical for the energy transition and meeting global climate targets, as they provide a systematic framework to evaluate the impact of natural phenomena, technological innovations, policy mandates, and socioeconomic factors. The variants of CEMs in the energy planning context include generation expansion (GEP), transmission expansion (TEP), and the combined generation and transmission expansion planning model (GTEP). Different versions of these models are defined by their problem parameters (e.g., spatio-temporal resolution), key decision-making entities, and major assumptions that govern their dynamics.  

Infrastructure planning is inherently forward-looking; hence, uncertainty is an intrinsic feature of the decision-making process. Key sources of uncertainty include weather and climate variability, technological advancements, policy and regulatory evolution, and societal factors \citep{FarrokhifarEtal2020_survey, RoaldEtal2023_survey}. These uncertainties can fundamentally alter both optimal investment strategies and operational outcomes. Despite this, a substantial portion of the CEM literature assumes perfect knowledge and resorts to deterministic formulations in which point estimates replace the value of uncertain parameters. This simplification has traditionally been justified by three considerations. First, detailed probabilistic information or an ensemble of plausible projections is often unavailable. Second, the most likely parameter values that are obtained by more accurate prediction models can yield sufficiently acceptable outcomes \citep{RoaldEtal2023_survey}. Third and most importantly, deterministic models are more computationally tractable, enabling extensive sensitivity analysis and large-scale applications within reasonable time frames. 
 
However, the global push toward decarbonization has amplified the limitations of deterministic planning. The mass adoption of variable renewable energy (VRE) such as solar and wind, electrification of end-use such as electric vehicles and heat pumps, and increasing interdependence between energy vectors have altered endogenous and exogenous uncertainties faced by planners. In particular, high penetration of VRE introduces intermittency, while electrification and flexibility resources such as battery storage add investment uncertainties related to cost trajectory, technology maturity, and performance capability. These dynamics reduce the practical relevance of deterministic models in favor of data-driven stochastic formulations. In parallel, recent methodological advances have expanded the tools available to explicitly incorporate uncertainty into CEMs. Modeling paradigms such as stochastic programming (SP), robust optimization (RO), and distributionally robust optimization (DRO) offer mathematical foundations to incorporate variability directly into the optimization model. These approaches allow planners to quantify trade-offs between cost, reliability, and risk.

\begin{table}[]
\scriptsize
    \centering
        \caption{Initialisms and acronyms used throughout the paper}
    \begin{tabular}{ll|ll}
    \toprule
   (D)RO& (Distributionally) robust optimization & SP& Stochastic program\\
    LOLE& Loss of load expectation & (C)VaR & Conditional value-at-risk \\
        2SP/mSP& Two/multi-stage stochastic program & VRE & Variable renewable energy \\
        2RO/mRO& Two/multi-stage robust optimization & CC& Chance constraint\\
        FOM/VOM& Fix/variable operating and maintenance cost& CEM& Capacity expansion models\\
          CCG& Cut-and-column generation algorithm & (MI)LP& (Mixed-integer) linear program\\
         MISOCP& Mixed-integer second order cone program& LDR & Linear decision rule \\
         BD & Benders decomposition algorithm& EF& Extensive form\\
         GEP/TEP & Generation/transmission expansion problem &EUE& Expected unserved energy\\
         GTEP& Generation and transmission expansion problem&RPS& Renewable portfolio standard\\
         mDRO/dDRO& DRO with moment/distance-based ambiguity set& DW& Dantzig-Wolfe Decomposition \\
         DER & Distributed energy resources &PHA &Progressive hedging algorithm\\
         SDDP& Stochastic dual dynamic programming &ISO& Independent system operator\\
         SA & Sensitivity analysis & MCS & Monte Carlo sampling\\
         CF & Capacity factor & UC & Unit commitment\\
         \bottomrule
    \end{tabular}
    \label{tab:abbs} 
    \vspace{-1em}
\end{table}

\begin{figure}[!t]
    \centering
    \includegraphics[width=\linewidth]{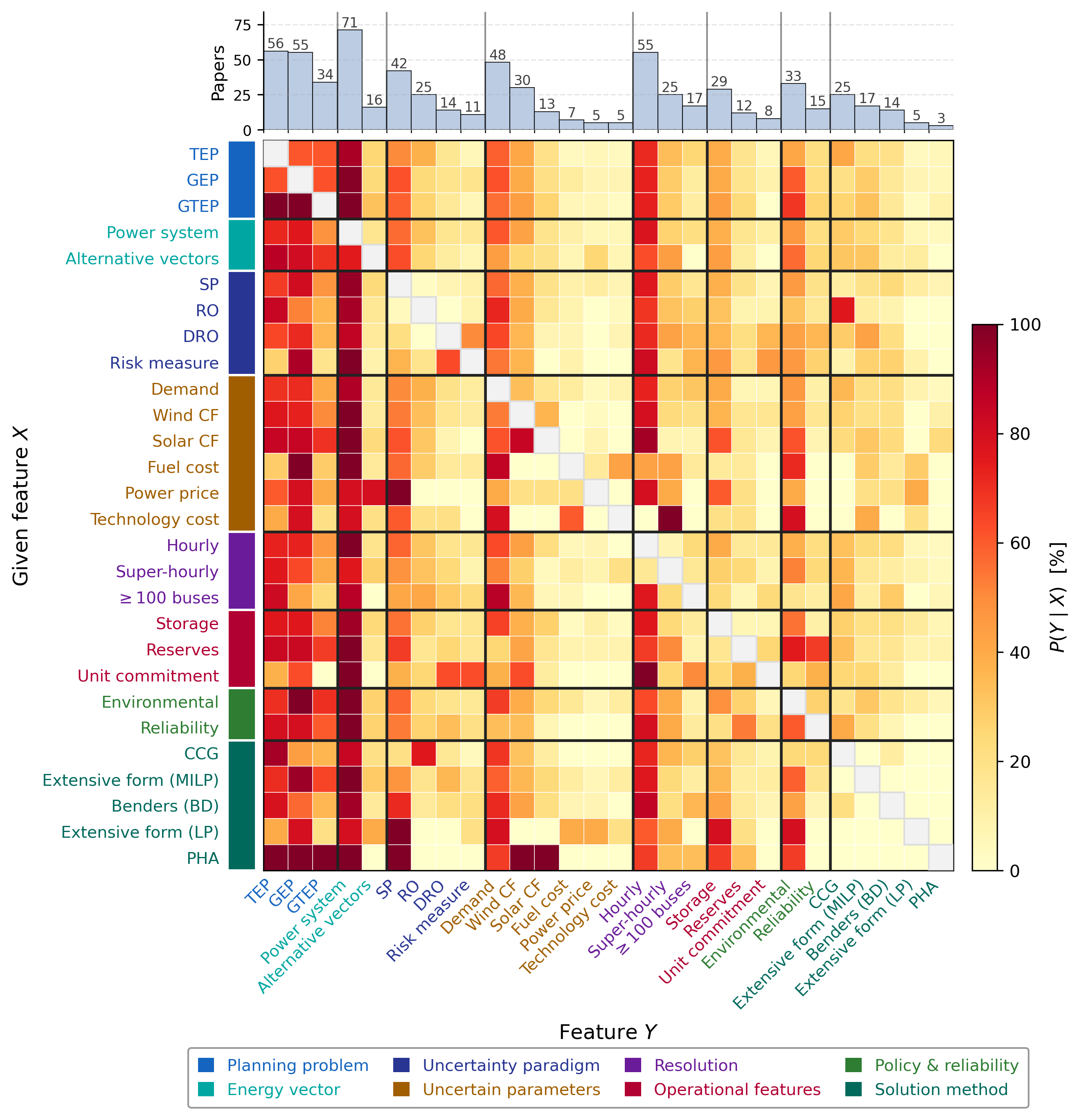}
    \caption{Conditional co-occurrence of modeling features across the surveyed papers. Cell $(X, Y)$ reports the percentage of papers containing feature $X$ that also contain feature $Y$, i.e., $P(Y \mid X)$; the matrix is therefore asymmetric. Rows and columns share the same ordering: features are grouped into the eight categories shown in the legend (colored band, left) and sorted by frequency within each group. Bars above the matrix give the number of papers containing each feature. Features appearing in fewer than three papers are omitted.}
    \label{fig:co-occurrence}
    \vspace{-1em}
\end{figure}

In this survey, we review recent advances at the intersection of energy infrastructure planning (modeled as GEP, TEP, and GTEP) and optimization under uncertainty (using SP, RO and DRO techniques). We restrict our literature review to scholarly articles published since 2015 to capture the latest developments, with particular emphasis on contributions that advance modeling formulations, uncertainty representations, and solution methodologies.
The literature was identified primarily through a systematic search in major bibliographic databases using a combination of keywords related to generation/transmission expansion, stochastic programming, robust optimization, and distributionally robust optimization. To ensure academic rigor and relevance, we focus mainly on peer-reviewed journal articles published in leading venues in operations research, electric power, and energy systems. While the resulting list of reviewed papers is not exhaustive, it provides representative coverage of the methodological developments that have shaped the recent literature. 
In addition, we review and taxonomize a selection of existing open-source capacity expansion models to provide a practical, tooling-based perspective that complements our literature review.

This survey complements prior reviews that have examined i) CEMs in power systems \citep{Koltsaklis2018_survey, GacituaEtal2018_survey_GTEP, Kaya2025_EJOR_survey} and integrated energy systems \citep{Klatzer2022_survey, FarrokhifarEtal2020_survey}; ii) optimization under uncertainty in power grid applications \citep{RoaldEtal2023_survey, Zakaria2020_survey}; or iii) specific themes in energy systems modeling, such as general modeling challenges \citep{Fodstad2022_survey}, the role of storage \citep{Levin2023_survey}, and incorporation of climate uncertainty \citep{Plaga2023_survey}. {This survey differs by using the intersection among planning decisions, uncertainty treatment, modeling fidelity, and practical implementation. Specifically, it offers four contributions. Specifically, it synthesizes how different sources and timescales of uncertainty are represented through different modeling paradigms across energy planning problems. It then relates these formulations to modeling fidelity and solution methods and derives the conditional co-occurrence analysis summarized in Fig.~\ref{fig:co-occurrence} to identify dominant combinations and underexplored intersections in the literature. Moreover, it compares representative open-source CEM tools to assess how methodological advances are translated into accessible and reproducible implementations and to identify capabilities that remain limited in existing software. Finally, the survey presents an interpretation of the findings relative to the needs of different decision-makers and stakeholders. } 

The paper is organized to first lay out the decision context and modeling purpose of energy infrastructure planning in Section~\ref{sec:decision-context}. Scope and preliminaries are given in Section~\ref{sec:scope_background}. Section~\ref{sec:review} presents the detailed survey of the reviewed papers. Remaining challenges and the potential research gaps are explained in Section~\ref{sec:gaps}. The emerging role of machine learning on energy infrastructure planning under uncertainty is discussed in Section~\ref{sec:ml_role}. We summarize the gaps and discuss their importance and implications in Section~\ref{sec:discussion}. Finally, Section~\ref{sec:conclusion} concludes the survey.

\section{Decision Contexts and Modeling Purpose}\label{sec:decision-context}
Energy infrastructure planning models serve diverse decision-makers whose objectives and informational needs differ substantially. As a result, the type and level of detail required from these models vary significantly. \textbf{Strategic investors} such as utilities and private equity firms are primarily concerned with aggregate indicators (e.g., total investment cost and expected returns), and risk-adjusted metrics (e.g., internal rate of return and net present value). In many cases, these decisions depend more on the economic viability of a portfolio than precise operational fidelity. Accordingly, differences in modeling choices, such as hourly vs. monthly, may not materially impact the investment decisions if they lead to similar total costs. 
\textbf{System operators}, including independent system operators (ISOs) and transmission system operators (TSO), face a different set of requirements. Their primary concern is operational feasibility and reliability, so they require models capable of verifying operational constraints and producing a detailed dispatch schedule. Reliability metrics such as loss-of-load expectation (LOLE) and expected unserved energy (EUE) are central to their decision-making \citep{He2017_TPS_PG_GTEP}, and modeling choices related to temporal resolution and operational constraints such as unit commitment (UC) and reserve requirements can directly impact resource adequacy and market design. \textbf{Regulators and public planners} such as state public utility commissions and national energy agencies tend to focus on broader societal and distributional outcomes. Their interests often include the allocation of costs across consumer classes, the equity implications of infrastructure siting, and the resilience of energy systems to extreme events.  For them, the importance of spatial granularity and the ability to evaluate distributional impact can exceed the operational details. Finally, \textbf{policymakers}, including legislative bodies and environmental agencies, typically approach planning tools to evaluate alternative policy pathways. They compare the performance of different policy instruments, such as the carbon tax and renewable portfolio standard (RPS), to understand broader economic and environmental implications. In this context, sensitivity analysis to highlight trade-offs among costs, emissions, and infrastructure outcomes is more valuable than engineering details. 

These different decision contexts indicate an important principle: modeling fidelity is valuable only relative to the decisions the model informs. 
A high-fidelity model that captures hourly UC but produces similar total cost estimates from a stylized model may only add computational burden without decision value for investors. Conversely, for system operators responsible for ensuring reliability, such operational details are indispensable. Throughout this survey, references to ``gaps'' in modeling fidelity are therefore not a blanket call to increase spatio-temporal resolution or incorporate more constraints. Rather, they reflect areas where the value of additional modeling detail relative to improvements in decision quality may be uncertain or insufficiently studied, but likewise potentially consequential, and thus require further assessment and/or new methodological innovation.

\section{Scope and Background}\label{sec:scope_background}
This section defines the scope of the survey. We define the classes of energy infrastructure considered, summarize the main sources of uncertainty that shape long-term planning problems, and outline the methodological foundations used to incorporate uncertainty into capacity expansion models. 

\subsection{Energy Infrastructure Planning}\label{ssec:infra-planning}
In modern societies, energy infrastructure comprises the physical assets that enable the reliable operation of essential services such as communication, transportation, lighting, and heating/cooling. While electricity and natural gas are central to most national energy systems, hydrogen and other low-carbon carriers are also expected to play an increasing role in future systems \citep{Klatzer2022_survey, Brown_H2_2023}. From an optimization perspective, hydrogen introduces additional decision layers including production, storage, and transport, each with distinct cost structures and coupling structures. These features necessitate optimization frameworks beyond the traditional power-centric formulations. Consequently, expanding and coordinating these assets is critical for supporting the energy transition, meeting growing global demand and improving resilience of energy systems.

To guide such expansions, planning models such as GEPs, TEPs, and  GTEPs are widely employed by national and regional planning authorities, system operators,  utilities,  generation companies, market regulators, energy consultants, and research institutions \citep{Levin2023_survey}. These models typically determine investment decisions that minimize the total system cost, including asset costs (new assets and enhancements of existing assets), operational costs, and environmental costs, subject to operational and physical constraints that ensure feasibility and reliability. 
Investment decisions, such as upgrading and/or retiring assets, are made on long-term horizons (years) and often across multiple planning periods. Given these investments, operational decisions then determine how to reliably balance supply and demand on shorter time scales. 

\textbf{GEPs} determine the optimal mix of new power generation capacity and storage facility, their type, location, and timing. 
\textbf{TEPs} address where and when to build or upgrade transmission lines, and by analogy gas or pipeline, so that generated electricity can be delivered to loads at acceptable cost and reliability. Finally, \textbf{GTEPs} integrate both sets of decisions and explicitly model the interdependencies between generation placement and power delivery. \citet{Kaya2025_EJOR_survey} {provide} a comprehensive survey on the optimization models in power systems, including the mathematical models for these problem classes. In this survey, we use `CEMs' as an umbrella term for all these problems, and extend the terminology as needed to describe similar decisions in gas and hydrogen systems. Therefore, in a gas system, for example, we use GEP and TEP to refer to the expansion of gas supplier assets and pipelines, respectively. 

{The CEMs adopt a centralized structure when a central planner coordinates all investment and operational decisions to minimize total system cost or maximize social welfare. This structure is widely used to provide a benchmark for system-wide planning, but it may not fully represent restructured markets where investment decisions are distributed among transmission planners (or operators), ISOs, and private firms with potentially different objectives and sets of constraints. Market-based and decentralized model represent try to emulate these interactions with game-theoretical or decentralized optimization approaches \citep{Wogrin2021_GTEP, Garcia2025_EJOR_GEP}. }

The feasible space of CEMs is defined by a core suite of constraints that reflect physical laws, operational limits, and modeling assumptions. {Although their implementation varies across applications, most formulations contain three groups of constraints}:
\begin{itemize}[leftmargin=1em, itemsep=3pt, parsep=0pt, topsep=0pt]
    \item \textbf{{System balance and resource limit}:} {Energy balance constraints} ensure that energy demand {is met by production, storage discharge and net import. Depending on the model, load shedding and curtailment may represent shortage and excess supply. Resource limits restrict production, or supply according to installed capacity and resource availability. For VRE, for example, output is bounded by installed capacity and time-varying capacity factors (CFs)}, which are parameters that indicate the ratio of actual power generation to the maximum generation potential over a period of time.
    \item {\textbf{Network flow constraints:}} {Network-aware CEMs represent the transfer or energy among supply, demand, and storage locations.} In power systems, the underlying alternating current (AC) power flow equations are {commonly approximated using} direct current (DC) power flow {or transportation representation to maintain tractability. Similarly,} in gas systems, nonlinear flow equations \citep{Wolf2000_MS_gas} are often simplified using transportation approximations. 
    \item {\textbf{Investment and expansion constraints:}} {These constraints govern the type, location, timing, and scale of infrastructure addition, upgrade, and retirement. They may reflect availability of candidate projects, construction lead time, land-use or siting restrictions, technology-specific limits, and total investment budget. }
\end{itemize}

Advances in computational resources and modeling techniques have allowed CEMs to incorporate richer operational details, making them increasingly complex.{These constraints can be organized into four broad groups:}

\begin{itemize}[leftmargin=1em, itemsep=3pt, parsep=0pt, topsep=0pt]
\item {\textbf{Resource operations and flexibility:}} {Thermal generator constraints include unit commitment (UC), ramping, and minimum up- and down-times. Storage constraints track the state of charge and impose energy capacity, charging, discharging, efficiency, and degradation limits \citep{Sioshansi2021_survey_storage}. Demand flexibility may also be modeled by allowing a portion of consumption to shift across periods under amount and shifting duration limits \citep{Motta2024_survey_DRP}.} 
\item {\textbf{Reliability and resilience requirements:}} {CEMs may ensure resource adequacy using metrics such as} loss-of-load expectation (LOLE) which measures the expected number of periods with unserved load, and expected unserved energy (EUE) {which quantifies its expected magnitude. Reliability can also be enforced through operating and planning reserves, limits on lost load, and $N-k$ contingency criteria requiring the system to withstand the simultaneous failure of $k$ components} (generator or transmission line). 
\item {\textbf{Policy constraints and instruments:}} {Policy considerations include renewable portfolio standards (RPS), renewable energy target (RET), carbon tax, emissions limit, and restrictions or incentives affecting the deployment of particular technologies. RPS and RET require minimum renewable generation requirements, whereas carbon taxes and emissions limits discourage generation from carbon-intensive resources \citep{Nunes2018_PG_GTEP}}. 
\item {\textbf{Cross-vector coupling constraints:}} {Integrated electricity, gas, and hydrogen models represent the physical and operational relationships among energy carriers. These constraints may capture natural-gas consumption by power generators, electricity consumption in gas and hydrogen infrastructure, conversion processes such as electrolysis, and joint emissions limits across multiple energy vectors.}
\end{itemize}

\subsection{Sources of Uncertainty and Their Evolution}
Energy infrastructure planning is fundamentally a problem of decision-making under uncertainty due to lack of complete knowledge of how key drivers will evolve. {The conventional expansion planning literature has focused primarily on factors that directly impact required capacity and investment expenditures, including demand growth, fuel prices and the availability and cost of generation resources \citep{FarrokhifarEtal2020_survey}. However, the sources of uncertainty have broadened significantly as energy systems have become increasingly dependent on VRE, electrification, interconnected energy vectors, and evolving decarbonization policies. Accordingly, recent models consider a diverse set of uncertain parameters such as wind and solar generation, hydro inflow, infrastructure availability, market prices, and weather and climate conditions \citep{RoaldEtal2023_survey, Plaga2023_survey}}. 

{One useful distinction concerns the unfolding of uncertainty over time. Long-term uncertainty primarily affects investment decisions as they unfold over years or decades. It stems from long-term demand growth, capital and fuel cost trajectories, technology learning and lifetime, and construction lead time. These factors impact the timing, location, type, and scale of infrastructure development. In contrast, short-term uncertainty captures variability over operational timescale, ranging from minutes to hours and seasons. It influences the utilization or performance of infrastructure after they become operational, and can include hourly demand and VRE generation, component outage, and short-term market prices. This distinction, however, is not absolute because some parameters affect the problem on multiple levels. Demand, for example, has direct impact on short-term operational decisions while its anticipated long-term projection affects generation, transmission, and storage investments. }

{Uncertainties may also be exogenous or endogenous. The former evolves independently of the planner's decisions and can include weather conditions and geopolitical disruptions. Endogenous or decision-dependent uncertainties arise when part of the decisions influence the realization, timing, or distribution of uncertain parameters \citep{Apap2017_DD_uncertainty}. Examples include electricity prices affected by installed generation capacity, electrification influenced by infrastructure availability, and demand response participation affected by market and network design. It is worth noting that at a broader level, structural shifts (e.g., abrupt regulatory changes, technological breakthroughs, and nonstationary climate patterns) may invalidate the assumptions or relationships embedded in a planning model. }

\subsection{Optimization under Uncertainty}
To address uncertainty, a broad set of stochastic optimization paradigms has been developed to incorporate uncertainty directly into CEMs. The three dominant frameworks are SP, RO, and DRO. 
These methods are valued for their flexibility in representing uncertainty while often admitting tractable reformulations \citep{RoaldEtal2023_survey}. Stochastic program is a popular method that assumes perfect knowledge of distributions governing the stochastic parameters. In practice, uncertainty is represented by a finite set of discrete scenarios, and the model optimizes the expected value of costs across these realizations \citep{Birge_book2011_SP}. {In its basic two-stage form, investment decisions are made before uncertainty is realized, while operational decisions adapt afterward. Multistage SP extends this to multiple decision epochs represented by a scenario tree, with nonanticipativity constraints ensuring that decisions depend only on the information available at each state \citep{Ding2017_TPS_PG_GTEP}. More advanced variants introduce intermediate levels of adaptability \citep{Basciftci2024_MSOM_GEP} or relax the assumption that uncertainty is exogenous, independent of the investment decisions \citep{Zhan2016_TPS_GEP}. }

Robust optimization addresses situations where the underlying distribution cannot be reliably estimated due to insufficient or noisy data, {or when protection against adverse realizations is more important than their probability of occurrence}. Instead of \textit{scenarios}, RO defines an \textit{uncertainty set},  often a bounded convex set, within which all realizations are considered plausible {\citep{Bertsimas2011}}. The model then safeguards against the worst-case realization(s) in the uncertainty set. Commonly used sets include box (i.e., interval), budget, ellipsoidal, and polyhedral \citep{RoaldEtal2023_survey, Sun2021_book_RO}. 
Adaptive or two-stage RO allows limited recourse actions after uncertainty realization, whereas multistage RO determines investment decisions across multiple time epochs. 

Distributionally robust optimization {relaxes the assumption that the probability distribution is known exactly. Instead, it optimizes the worst-case distribution within an \textit{ambiguity set} constructed from available data or statistical information \citep{RahimianMehrotra2019_DRO_review}}. 
The moment-based ambiguity sets contain distributions with similar moments (mean, covariance) or other statistical measures such as modality {\citep{Delage2010_DRO}}. Distance-based ambiguity sets identify those distributions whose probabilistic distance from the nominal empirical distribution is restricted to a given positive value \citep{MohajerinKuhn2018_MathProg}. {More advanced DRO models incorporate additional distributional structure, such as modality information, and combine ambiguity st with risk measures such as CVaR \citep{Pourahmadi2019_TPS_GEP, Pourahmadi2021_TPS_GEP}. Other extensions represent uncertainty at multiple scales by combining long-term scenario with conditional ambiguity sets for short-term operational uncertainty \citep{Velloso2020_TPS_TEP}, or allowing planning decisions to affect the probability of uncertain parameters \citep{Alvarado2022_TPS_TEP}.}.

Each framework also admits extensions that incorporate explicit measures of \textit{risk} to capture the preference of decision makers. Risk-averse models employ metrics such as value-at-risk (VaR),  conditional value-at-risk (CVaR), shortfall risk, and chance constraint (CC) to manage downside exposure and to reflect the asymmetric consequences of extreme events. Such extensions are particularly relevant in long-term infrastructure planning in which asymmetry in the consequences of uncertain events is often more important than their expected frequency. Comprehensive reviews of these approaches and associated risk measures {are provided by} \citet{RoaldEtal2023_survey, Zakaria2020_survey, Geng2019_CC_survey}.

Each approach presents distinct challenges and advantages. SP requires information-rich data to estimate the true distributions, but this can hamper the problem scalability \citep{BertsimasEtal2023_MSOM}. Furthermore, in the energy planning context, even abundant empirical data may fail to capture structural changes, such as those induced by climate change, hence undermining assumptions about \textit{future} distributions. Overfitting to sampled scenarios is another recognized drawback, as decisions optimized on a limited dataset may perform poorly on unseen realizations \citep{VanParys2021_MS}. 
RO, on the other hand, is motivated by data scarcity or corruption, yet it typically ignores structural properties of the data and can lead to overly conservative results \citep{Goerigk2023_COR_RO_NN}. DRO offers a compromise by systematically incorporating partial statistical information and controlling the degree of conservatism, but this often comes at the cost of increased modeling and computational complexity. Except for certain ambiguity sets (e.g., first moment, L1-norm) that are amenable to tractable reformulations, DRO problems can lead to nonlinear, nonconvex, or even infinite-dimensional formulations \citep{Sun2021_book_RO}. 

{It is worth noting that a complementary framework for organizing the uncertainty modeling paradigms is Powell's universal framework for sequential decision under uncertainty \citep{Powell2016UnifiedFramework, Powell2022ReinforcementLearning}. It represents a sequential decision problem through five elements: state variable, decision variables, exogenous information, a transition function, and an objective function. In a multistage CEM, for instance, the state may include installed infrastructure and currently available information; decisions are investment and operational variables; exogenous information include problem parameters (e.g., demand and costs); and the transition function describes how decisions and new information change the system over time. The central object is therefore a \textit{policy} that maps the current state to a feasible action, rather than a fixed sequence of future decisions.}
 
\normalsize
\section{Energy Infrastructure Expansion Models}\label{sec:review}
In this section, we review recent contributions to energy infrastructure planning under uncertainty as well as open-source CEM tools. For each topic, we summarize the literature in a table to provide a structured overview. In all tables, the column ``Solution approach'' specifies the general methodology used for solving instances. Some papers reformulate the problem as a large-scale optimization model and solve directly in its extensive form (EF), typically as a linear program (LP), mixed-integer linear program (MILP), mixed-integer second-order cone program (MISOCP), or quadratic program (QP). Other studies adopt solution strategies inspired by well-known decomposition paradigms such as the progressive hedging algorithm (PHA), cut-and-column generation (CCG), Benders decomposition (BD), alternating direction method of multipliers (ADMM), or Dantzig-Wolfe (DW) decomposition. 

The column ``Granularity'' indicates the spatio-temporal resolution of the largest instances considered in the corresponding paper. On the left side of the $|$ symbol, we report the temporal resolution, number of stages, and number of scenarios (in SPs). On the right-hand side, {we report the number of spatial units (nodes, buses, zones); a value of one denotes a copperplate or single-bus representation.} When available, we also include the planning horizon per scenario. For instance, \textit{`1h, 4 stages, 8
5-yr scen.| 1'} denotes that the largest instance has an hourly resolution (1h), is a multistage SP formulated on a scenario tree with 4 stages and 8 scenarios, each scenario spanning 5 years (5$\times$ 8760 hours), and is implemented as a single bus model (1 node). 
Finally, the column ``Key aspects'' describes the distinctive modeling assumptions or methodological contributions of each paper. Where necessary, we provide additional descriptions of modeling details in the text. Each section also discusses the main cost components and system constraints considered across the literature.

\subsection{Power Generation Expansion Problems}
The overview of GEPs under uncertainty summarized in Table~\ref{tab:GEP} shows {both persistence and diversification in the optimization paradigms}.
\citet{Thangavelu2015_AE_GEP} ensure diversity of generation mix in the outcomes as a proxy for energy security. \citet{Park2015_TPS_GEP} generate scenarios by the Gaussian copula method applied to the limited realizations. \citet{Zhan2016_TPS_GEP} {propose a decision-dependent SP}, where the distribution of electricity prices depends on the installed generators' capacity. 
\citet{Costa2017_EE_GEP} formulate the problem as a portfolio optimization with uncertain returns and a covariance matrix subject to lower and upper bounds of each portfolio component (i.e., generation capacity).

{Six of the seven studies published before 2020 use SP. Among the papers published after 2020, while SP remains the most common paradigm, there is a greater adoption of DRO. The choice of paradigm is generally consistent with the information available about uncertainty and the desired form of hedging. SP is commonly used when uncertainty can be represented through discrete scenarios, and RO and DRO are preferred when reliable probability distribution is unavailable}.
\citet{Pourahmadi2019_TPS_GEP} and \citet{Pourahmadi2021_TPS_GEP} use moment-based DRO to capture short-term wind uncertainty, while relying on SP for long-term demand trends. \citet{DaCosta2020_TPS_GEP} {propose a} multistage SP model with reliability measures such as VaR and CVaR imposed as constraints. \citet{Pourahmadi2021_TPS_GEP} and \citet{Hu2022_EPSR_GEP} {impose} modality information of nominal distribution as part of the ambiguity set. \citet{Abdin2022_AE_GEP} consider a budgeted polyhedral uncertainty set for a multistage RO. \citet{Guevara2020_AE_GEP} and \citet{Chen2022_TPS_GEP} {use a} DRO with Wasserstein distance defined on the first norm. As a middle ground between multi- and two-stage SP, \citet{Basciftci2024_MSOM_GEP} {present} an adaptive approach, in which it considers a revision stage in the scenario tree for each investment decision to consolidate all the previous decisions in the previous stages. Finally, \citet{Garcia2025_EJOR_GEP} develop a bilevel model where the leader is a strategic investor that determines the capacity expansion to maximize its profit, and the follower is an ISO that maximizes social welfare for all market participants.  

{Uncertain parameters are limited to small set of parameters.}
The most frequently modeled uncertain parameters are electricity demand and wind generation, followed by solar generation. \citet{Zhan2016_TPS_GEP} {model electricity price uncertainty} as a multi-stage SP. 
\citet{Moret2020_EJOR} consider uncertainty in technology lifetime, investment costs, FOM, discount rate, demand, supply, and conversion efficiencies. In the suquel paper, \citet{Guevara2020_AE_GEP} apply a decision-tree method 
to identify parameters whose uncertainty has the largest impact on the problem outcome. These parameters are identified as import costs for natural gas, electricity, and coal, as well as heat rates of some thermal generators. \citet{Garcia2025_EJOR_GEP} {consider} the uncertainty of peak intercept of the linear inverse-demand curve (i.e., value of lost load) and marginal production cost.

\begin{table}[!b]
\small 
\caption{Overview of generation expansion problems under uncertainty for power system}
\vspace{-0.2cm}
{\scriptsize
\setlength{\tabcolsep}{7pt}
\renewcommand{\arraystretch}{1.1} 
\begin{tabular}{l|llllllll}
\toprule
\multirow{2}{0.08\columnwidth}{Reference} &
\multirow{2}{0.05\columnwidth}{Model}&  \multirow{2}{0.07\columnwidth}{Uncertain parameters}& \multirow{2}{0.05\columnwidth}{Solution, Model}&\multirow{2}{0.12\columnwidth}{Granularity}&\multirow{2}{0.35\columnwidth}{Key aspects}\\ \\
\midrule
 \multirow{3}{0.13\columnwidth}{\citet{Thangavelu2015_AE_GEP}}&\multirow{3}{0.055\columnwidth}{mSP}& \multirow{3}{0.11\columnwidth}{demand, fuel/tech cost}& \multirow{3}{0.05\columnwidth}{EF, NLP}  &\multirow{3}{0.12\columnwidth}{1yr,~5 stages, 57 scen. |1} &\multirow{3}{0.35\columnwidth}{RPS, emissions limit, carbon tax, fuel diversity constraint} \\ \\ \\
\midrule
 \multirow{2}{0.13\columnwidth}{\citet{Park2015_TPS_GEP}}&\multirow{2}{0.055\columnwidth}{2SP}& \multirow{2}{0.1\columnwidth}{demand, wind gen.}& \multirow{2}{0.05\columnwidth}{BD, {MILP}}  &\multirow{2}{0.12\columnwidth}{1h, 300 1h scen. |1} &\multirow{2}{0.35\columnwidth}{RPS, carbon tax} \\ \\ 
 \midrule
 \multirow{2}{0.13\columnwidth}{\citet{Zhan2016_TPS_GEP}}&\multirow{2}{0.055\columnwidth}{mSP}& \multirow{2}{0.1\columnwidth}{electricity price}& \multirow{2}{0.05\columnwidth}{EF, MILP}  &\multirow{2}{0.12\columnwidth}{1h, 4 stages, 8 5-yr scen. |1} & \multirow{2}{0.35\columnwidth}{price distribution depends on installed capacity of generators} \\ \\ 
  \midrule
\multirow{2}{0.13\columnwidth}{\citet{Kocaman2017_AE_GEP}}&\multirow{2}{0.055\columnwidth}{2SP}& \multirow{2}{0.1\columnwidth}{solar gen., dem., inflow}& \multirow{2}{0.05\columnwidth}{EF, MILP}  &\multirow{2}{0.12\columnwidth}{3h, 13 yearly scen. |4} & \multirow{2}{0.35\columnwidth}{the expansion of pumped hydro storage is considered} \\ \\ 
  \midrule
\multirow{2}{0.13\columnwidth}{\citet{Costa2017_EE_GEP}}&\multirow{2}{0.055\columnwidth}{2RO}& \multirow{2}{0.11\columnwidth}{fuel, VOM, emiss. costs}& \multirow{2}{0.05\columnwidth}{EF, SOCP}  &\multirow{2}{0.12\columnwidth}{1yr, 1yr |1} & \multirow{2}{0.35\columnwidth}{GEP as a portfolio optimization} \\ \\ 
  \midrule
\multirow{3}{0.13\columnwidth}{\citet{Ioannou2019_EE_GEP}}&\multirow{3}{0.055\columnwidth}{mSP}& \multirow{3}{0.11\columnwidth}{demand, fuel/tech cost}& \multirow{3}{0.05\columnwidth}{EF, LP}  &\multirow{3}{0.12\columnwidth}{1yr, 3 stages, {1.09e5} scen. |1} & \multirow{3}{0.35\columnwidth}{RPS, emissions limit, carbon tax} \\ \\ \\
  \midrule
\multirow{2}{0.13\columnwidth}{\citet{Pourahmadi2019_TPS_GEP}}&\multirow{2}{0.055\columnwidth}{mDRO, 2SP}& \multirow{2}{0.1\columnwidth}{wind gen., demand}& \multirow{2}{0.05\columnwidth}{EF, MISOCP}  &\multirow{2}{0.12\columnwidth}{1h,~1~yr, 10 rep. days |118} & \multirow{2}{0.35\columnwidth}{LDR and convexification of UC constraints} \\ \\ 
  \midrule
\multirow{2}{0.13\columnwidth}{\citet{Moret2020_EJOR}}&\multirow{2}{0.05\columnwidth}{RO} & \multirow{2}{0.1\columnwidth}{all param.}& \multirow{2}{0.05\columnwidth}{EF, MILP}  &\multirow{2}{0.12\columnwidth}{1month,~1~yr, |1} & \multirow{2}{0.35\columnwidth}{storage considered, transmission loss} \\ \\ 
  \midrule
\multirow{2}{0.13\columnwidth}{\citet{Guevara2020_AE_GEP}}&\multirow{2}{0.05\columnwidth}{dDRO} & \multirow{2}{0.1\columnwidth}{import cost, heat rates}& \multirow{2}{0.05\columnwidth}{BD, {MILP}}  &\multirow{2}{0.12\columnwidth}{1month,~1~yr, |1} & \multirow{2}{0.35\columnwidth}{storage considered, ML identifies main uncertain parameters, transmission loss} \\ \\ 
  \midrule
\multirow{2}{0.13\columnwidth}{\citet{DaCosta2020_TPS_GEP}}&\multirow{2}{0.05\columnwidth}{mSP} & \multirow{2}{0.1\columnwidth}{wind gen.}& \multirow{2}{0.05\columnwidth}{SDDP, {MILP}}  &\multirow{2}{0.12\columnwidth}{1h, 7 stages |1} & \multirow{2}{0.35\columnwidth}{partial load flexibility, reliability-constrained formulation} \\ \\ 
  \midrule
\multirow{3}{0.15\columnwidth}{\citet{Pourahmadi2021_TPS_GEP}}&\multirow{3}{0.055\columnwidth}{mDRO, 2SP} & \multirow{3}{0.1\columnwidth}{wind gen., demand}& \multirow{3}{0.05\columnwidth}{EF, MISOCP}  &\multirow{3}{0.12\columnwidth}{1h,~1~yr, 10 rep. days |118} & \multirow{3}{0.35\columnwidth}{distributionally robust CC and CVaR, LDR and convexification of UC constraints 
} \\ \\ \\
  \midrule 
\multirow{2}{0.15\columnwidth}{\citet{Abdin2022_AE_GEP}}&\multirow{2}{0.05\columnwidth}{mRO} & \multirow{2}{0.1\columnwidth}{demand, VRE gen.}& \multirow{2}{0.05\columnwidth}{EF, MILP}  &\multirow{2}{0.12\columnwidth}{1h,~3~stages, 4 rep. days |12} & \multirow{2}{0.35\columnwidth}{RPS, apply LDR to reformulate as MILP} \\ \\ 
  \midrule
\multirow{2}{0.15\columnwidth}{\citet{Hu2022_EPSR_GEP}}&\multirow{2}{0.05\columnwidth}{mDRO} & \multirow{2}{0.1\columnwidth}{wind gen.}& \multirow{2}{0.05\columnwidth}{EF, MISOCP}  &\multirow{2}{0.12\columnwidth}{1h,~1~yr, 8 rep. days |118} & \multirow{2}{0.35\columnwidth}{RPS, distributionally robust CC, regulating reserve constraints} \\ \\ 
\midrule
\multirow{2}{0.15\columnwidth}{\citet{Chen2022_TPS_GEP}}&\multirow{2}{0.05\columnwidth}{dDRO} & \multirow{2}{0.1\columnwidth}{wind gen.}& \multirow{2}{0.05\columnwidth}{EF, MILP}  &\multirow{2}{0.12\columnwidth}{1h,~5~yrs, 1 rep. day |1354} & \multirow{2}{0.35\columnwidth}{distributionally robust CC, partial load flexibility, thermal storage considered} \\ \\ 
\midrule
\multirow{3}{0.15\columnwidth}{\citet{Irawan2023_EJOR_GEP}}&\multirow{3}{0.05\columnwidth}{2SP} & \multirow{3}{0.12\columnwidth}{total/peak demand, fuel cost}& \multirow{3}{0.05\columnwidth}{EF, MILP}  &\multirow{3}{0.12\columnwidth}{1h, 6 periods, 4000 scen.|6} & \multirow{3}{0.35\columnwidth}{carbon tax, transmission loss, goal programming to determine the energy mix} \\ \\ \\
\midrule
\multirow{2}{0.15\columnwidth}{\citet{Basciftci2024_MSOM_GEP}}&\multirow{2}{0.05\columnwidth}{adaptive 2SP} & \multirow{2}{0.12\columnwidth}{demand}& \multirow{2}{0.05\columnwidth}{{EF, MILP}}  &\multirow{2}{0.12\columnwidth}{1h, {8 stages}, 4 rep. days|1} & \multirow{2}{0.35\columnwidth}{approximation schemes for large-scale 2SP and mSP} \\ \\ 
\midrule
\multirow{3}{0.15\columnwidth}{\citet{Garcia2025_EJOR_GEP}}&\multirow{3}{0.05\columnwidth}{2SP, 2RO} & \multirow{3}{0.12\columnwidth}{lost load penalty, other}& \multirow{3}{0.05\columnwidth}{CCG, {MILP}}  &\multirow{3}{0.12\columnwidth}{1h,{10 rep. hours}, 100 scen.|6} & \multirow{3}{0.35\columnwidth}{CVaR, bilevel model} \\ \\ \\
\midrule
\multirow{2}{0.15\columnwidth}{\citet{Zampara2025_GEP}}&\multirow{2}{0.05\columnwidth}{2SP} & \multirow{2}{0.12\columnwidth}{demand, VRE gen., inflow}& \multirow{2}{0.07\columnwidth}{subgrad. method,{LP}}  &\multirow{2}{0.12\columnwidth}{1h, 35 1yr scen.|56} & \multirow{2}{0.35\columnwidth}{storage considered, EUE} \\ \\ 
\midrule
\multirow{2}{0.15\columnwidth}{\citet{Nygaard2025_GEP}}&\multirow{2}{0.05\columnwidth}{2SP} & \multirow{2}{0.1\columnwidth}{demand}& \multirow{2}{0.065\columnwidth}{EF, LP}  &\multirow{2}{0.12\columnwidth}{1h, 4 yearly scen.|3} & \multirow{2}{0.35\columnwidth}{emissions limit, storage considered, CVaR, carbon tax} \\ \\ 
\bottomrule
\end{tabular}}
\label{tab:GEP}
\end{table}

\normalsize

\subsection{Power Transmission Expansion Problems}
The overview of TEPs under uncertainty summarized in Table~\ref{tab:TEP} shows that most early studies use two-stage RO while more recent works adopt DRO and SP models. \citet{Velloso2020_TPS_TEP} model the long-term demand growth and VRE adoption as SP and short-term net load uncertainty by DRO. \citet{Alvarado2022_TPS_TEP} consider line hardening to enhance resiliency against earthquakes and model the pre-contingency preparations as SP and utilize the DRO framework to model the decision-dependent post-contingency failure probabilities of components (generators, lines, substations). The model includes DER services such as load increase/decrease via connection/disconnection of distributed generators to the grid as well as load shifting as a form of partial demand flexibility. Similarly, \citet{Alvarado2018_TPS_TEP} consider DER services and model the failure probability of generators and transmission lines as DRO with a tailored moment-based ambiguity set.  \citet{Giannelos2025_TEP} endogenously model the uncertainty of demand response programs to depend on line expansion decisions, and propose a framework to extend the scenario to represent the combined exogenous-endogenous uncertainties.
The most common uncertain variable is demand, followed by the VRE generation and the nameplate capacity of generators. The probabilistic nature of component failure is captured in \citep{Alvarado2018_TPS_TEP, Alvarado2022_TPS_TEP}. \citet{Zhang2017_TPS_TEP} and \citet{Giannelos2025_TEP} {model} peak demand uncertainty, which is central to reserve requirements; additionally, \citet{TapiaEtal2026_TEP}, {consider the uncertainty of VOM cost}. Finally, \citet{TapiaEtal2026_TEP} {consider} wildfire ignition risk and its impact on TEPs with two uncertainty sets define for line de-energization (disconnecting the line) as well as solar and wind generation.

\begin{table}[!b]
\small 
\caption{Overview of transmission expansion problems under uncertainty in power systems}
\vspace{-0.2cm}
{\scriptsize
\setlength{\tabcolsep}{7pt}
\renewcommand{\arraystretch}{1.1} 
\begin{tabular}{l|llllllll}
\toprule
\multirow{2}{0.08\columnwidth}{Reference} &
\multirow{2}{0.05\columnwidth}{Model}&  \multirow{2}{0.07\columnwidth}{Uncertain parameters}& \multirow{2}{0.05\columnwidth}{Solution, Model}&\multirow{2}{0.12\columnwidth}{Granularity}&\multirow{2}{0.35\columnwidth}{Key aspects}\\ \\
\midrule
 \multirow{2}{0.13\columnwidth}{\citet{Ruiz2015_EJOR_TEP}}&\multirow{2}{0.05\columnwidth}{2RO}& \multirow{2}{0.1\columnwidth}{demand, gen. cap.}& \multirow{2}{0.05\columnwidth}{CCG, {MILP}}  &\multirow{2}{0.12\columnwidth}{1h,~1~yr |24} &\multirow{2}{0.35\columnwidth}{{regional uncertainty budget}} \\ \\ 
 \midrule
\multirow{2}{0.13\columnwidth}{\citet{Dehghan2015_TSE_TEP}}&\multirow{2}{0.05\columnwidth}{2RO}& \multirow{2}{0.1\columnwidth}{demand, wind gen.}& \multirow{2}{0.05\columnwidth}{CCG, {MILP}}  &\multirow{2}{0.12\columnwidth}{1h,~1~yr, 4 rep. days |73} &\multirow{2}{0.35\columnwidth}{storage expansion, optimal transmission switching} \\ \\ 
 \midrule
 \multirow{2}{0.15\columnwidth}{\citet{Minguez2016_EJOR_TEP}}&\multirow{2}{0.05\columnwidth}{2RO}& \multirow{2}{0.1\columnwidth}{demand, gen. cap.}& \multirow{2}{0.05\columnwidth}{CCG, {MILP}}  &\multirow{2}{0.12\columnwidth}{1 yr,~1~yr |{118}} &\multirow{2}{0.35\columnwidth}{comparison of outcomes between previous related papers} \\ \\ 
  \midrule
 \multirow{2}{0.15\columnwidth}{\citet{Bagheri2016_TPS_TEP}}&\multirow{2}{0.05\columnwidth}{dDRO}& \multirow{2}{0.1\columnwidth}{demand}& \multirow{2}{0.05\columnwidth}{CCG, {MILP}}  &\multirow{2}{0.12\columnwidth}{time block,~20 yrs |118} &\multirow{2}{0.35\columnwidth}{{hybrid BD and CCG algorithm}} \\ \\ 
   \midrule
 \multirow{2}{0.15\columnwidth}{\citet{Garcia2016_TPS_TEP}}&\multirow{2}{0.05\columnwidth}{2RO}& \multirow{2}{0.11\columnwidth}{demand, gen. capacity}& \multirow{2}{0.05\columnwidth}{CCG, {MILP}}  &\multirow{2}{0.12\columnwidth}{1yr, 10 yrs |118} &\multirow{2}{0.35\columnwidth}{{}} \\ \\ 
  \midrule
 \multirow{2}{0.14\columnwidth}{\citet{Dehghan2017_TPS_TEP}}&\multirow{2}{0.05\columnwidth}{2RO}& \multirow{2}{0.1\columnwidth}{demand, wind gen.}& \multirow{2}{0.05\columnwidth}{EF, MILP}  &\multirow{2}{0.12\columnwidth}{1h,~1~yr, 20 rep. hrs |236} &\multirow{2}{0.35\columnwidth}{LDR to reformulate the problem as MILP} \\ \\ 
   \midrule
 \multirow{2}{0.14\columnwidth}{\citet{Zhang2017_TPS_TEP}}&\multirow{2}{0.05\columnwidth}{2RO}& \multirow{2}{0.12\columnwidth}{peak demand, gen. cap.}& \multirow{2}{0.05\columnwidth}{CCG, {MILP}}  &\multirow{2}{0.12\columnwidth}{1h,~1~yr, 45 rep. days |118} &\multirow{2}{0.35\columnwidth}{} \\ \\ 
   \midrule
 \multirow{2}{0.14\columnwidth}{\citet{Alvarado2018_TPS_TEP}}&\multirow{2}{0.05\columnwidth}{mDRO}& \multirow{2}{0.1\columnwidth}{component failure}& \multirow{2}{0.05\columnwidth}{BD, {MILP}}  &\multirow{2}{0.12\columnwidth}{0.5h,~1~yr, {40} rep. hrs |118} &\multirow{2}{0.35\columnwidth}{DER services, spinning reserve, $N-1$ contingency} \\ \\
    \midrule
 \multirow{2}{0.14\columnwidth}{\citet{Dehghan2018_TPS_TEP}}&\multirow{2}{0.05\columnwidth}{mRO}& \multirow{2}{0.1\columnwidth}{demand, wind gen.}& \multirow{2}{0.05\columnwidth}{EF, MILP}  &\multirow{2}{0.12\columnwidth}{1h, 5yrs, 5 rep. hrs |118} &\multirow{2}{0.35\columnwidth}{lift uncertainty sets to higher dimensions and apply binary decision rule and LDR} \\ \\
\midrule
\multirow{2}{0.14\columnwidth}{{\citet{Roldan2018_TEP}}}&\multirow{2}{0.05\columnwidth}{{2RO}}& \multirow{2}{0.1\columnwidth}{{demand, wind gen.}}& \multirow{2}{0.05\columnwidth}{{CCG, MILP}}  &\multirow{2}{0.12\columnwidth}{{1 rep. hr|2383}} &\multirow{2}{0.35\columnwidth}{{ellipsoid uncertainty set}} \\ \\
      \midrule
 \multirow{2}{0.14\columnwidth}{\citet{Zhan2018_TPS_TEP}}&\multirow{2}{0.05\columnwidth}{2SP}& \multirow{2}{0.1\columnwidth}{demand, wind gen.}& \multirow{2}{0.05\columnwidth}{BD, {MILP}}  &\multirow{2}{0.12\columnwidth}{1h,~3~yr, 15000 rep. hrs |300} &\multirow{2}{0.35\columnwidth}{installation of dynamic thermal rating devices for overhead lines} \\ \\ 
     \midrule
 \multirow{2}{0.14\columnwidth}{\citet{Zhuo2019_TPS_TEP}}&\multirow{2}{0.05\columnwidth}{2SP}& \multirow{2}{0.1\columnwidth}{demand, wind gen.}& \multirow{2}{0.05\columnwidth}{BD, {MILP}}  &\multirow{2}{0.12\columnwidth}{1h, 8760 scen.| 230} &\multirow{2}{0.35\columnwidth}{scenario clustering in each BD iteration for faster convergence} \\ \\
    \midrule
 \multirow{2}{0.14\columnwidth}{\citet{Wang2019_TPS_TEP}}&\multirow{2}{0.05\columnwidth}{2RO}& \multirow{2}{0.1\columnwidth}{VRE generation}& \multirow{2}{0.05\columnwidth}{CCG, {MILP}}  &\multirow{2}{0.12\columnwidth}{1h,~1~yr, (-) |196} &\multirow{2}{0.35\columnwidth}{storage expansion considered} \\ \\ 
\midrule
\multirow{2}{0.14\columnwidth}{\citet{Velloso2020_TPS_TEP}}&\multirow{2}{0.05\columnwidth}{2SP, mDRO}& \multirow{2}{0.1\columnwidth}{demand, wind gen.}& \multirow{2}{0.05\columnwidth}{CCG, {MILP}}  &\multirow{2}{0.12\columnwidth}{4h,~1~yr, 1 rep. day |118} &\multirow{2}{0.35\columnwidth}{CCG is enhanced by a DW procedure} \\ \\ 
\midrule
\multirow{2}{0.14\columnwidth}{\citet{Alvarado2022_TPS_TEP}}&\multirow{2}{0.05\columnwidth}{2SP, mDRO}& \multirow{2}{0.1\columnwidth}{component failure}& \multirow{2}{0.05\columnwidth}{CCG, {MILP}}  &\multirow{2}{0.12\columnwidth}{0.5h,~1 yr, 1 rep. hr |40} &\multirow{2}{0.35\columnwidth}{DER services, line hardening as a resiliency measure} \\ \\ 
\midrule
\multirow{2}{0.14\columnwidth}{\citet{Garcia2023_AE_TEP}}&\multirow{2}{0.05\columnwidth}{2RO}& \multirow{2}{0.14\columnwidth}{peak demand, gen. cap., VOM}& \multirow{2}{0.05\columnwidth}{CCG, {MILP}}  &\multirow{2}{0.12\columnwidth}{1h,~10 yrs, 80 rep. hrs |118} &\multirow{2}{0.35\columnwidth}{storage considered} \\ \\ 
\midrule
\multirow{3}{0.14\columnwidth}{\citet{Borozan2024_EPSR_TEP}}&\multirow{3}{0.05\columnwidth}{mSP}& \multirow{3}{0.13\columnwidth}{demand, VRE gen.}& \multirow{3}{0.05\columnwidth}{BD, {MILP}}  &\multirow{3}{0.12\columnwidth}{1h, 4 stages, 27 scen., 4 rep. days |118} &\multirow{3}{0.35\columnwidth}{storage expansion, supervised learning to identify promising optimality cuts} \\ \\ \\
\midrule
\multirow{2}{0.14\columnwidth}{\citet{Giannelos2025_TEP}}&\multirow{2}{0.05\columnwidth}{mSP}& \multirow{2}{0.13\columnwidth}{peak demand, other}& \multirow{2}{0.05\columnwidth}{BD, {MILP}}  &\multirow{2}{0.12\columnwidth}{1h, 5 stages, 24 scen.|123} &\multirow{2}{0.35\columnwidth}{endogenous uncertainty}\\ \\
\midrule
\multirow{2}{0.14\columnwidth}{{\citet{TapiaEtal2026_TEP}}}&\multirow{2}{0.05\columnwidth}{{2RO}}& \multirow{2}{0.13\columnwidth}{{VRE gen., other}}& \multirow{2}{0.05\columnwidth}{{CCG, MILP}}  &\multirow{2}{0.12\columnwidth}{{1h, 1yr, 3 rep. weeks|24}} &\multirow{2}{0.35\columnwidth}{{storage expansion  and underground line expansion considered}}\\ \\
\bottomrule
\end{tabular}}
\label{tab:TEP}
\end{table}

\normalsize

\subsection{Power Generation and Transmission Expansion Problems}
As presented in Table~\ref{tab:GTEP}, the literature of GTEP for power systems includes all modeling paradigms. {The table suggests that the choice between uncertainty modeling paradigms is determined less by its scope and impact than by the information available about it and the type of protection sought. SP favors full characterization of uncertainty, thus is a preferred option in modeling hydro-climatic pathways, demand and cost trajectories, and event-based representation of hurricane or component failure.}
Within SP-based approaches,  
\citet{Parkinson2015_GTEP} model the hydro-climatic trends by mid-century, considering pumped hydro storage and flexible demand amounting to up to 10\% of the peak demand. \citet{Pineda2016_EJOR_GTEP} model the wind generation forecast error as discrete scenario realizations and propose a bilevel model in which the balancing re-dispatch decisions for each scenario form the follower's problem. \citet{Liu2017_TPS_GTEP} consider the uncertainties in investment costs for wind and solar and the operational costs of coal and gas-fired power plants. \citet{Wogrin2021_GTEP} {propose} a multi-stage SP that recognizes the sequence of transmission expansion in which the lines are first developed and generators react by siting and sizing new generators. As such, they model this interplay as a stochastic bilevel program with the transmission expansion entity as the leader and generators as followers who engage in a Cournot competition among themselves. \citet{Bennett2021_GTEP} model the impact, severity, and probability of extreme hurricane events on Puerto Rico's grid via multi-stage SP. \citet{Zhang2024_COR_GTEP} {propose} a multi-horizon SP which is an approximation of a multistage SP and allows decoupling of decision epochs into long- and short-term stages for better tractability. The authors consider the demand and VRE generation as short-term, and CO$_2$ budget/tax and demand trends as long-term uncertainties. A resilient-informed multi-period two-stage SP model is proposed in \citep{Sahin2025_GTEP} in which the scenarios represent the set of failed substations. \citet{Rintamaki2023_TPS_GTEP} model the long-term uncertainty via multi-period RO and short-term uncertainties by representative days, each one treated as scenarios of the SP. Finally, \citet{LeeSun2025_GTEP} propose a solution methodology based on the level-bundle method to solve a contingency-aware model with high spatio-temporal resolution. 

{RO is preferred when the probability distribution is difficult to estimate, or hedging against adverse realization is desired. Therefore, RO is employed to model demand, VRE generation, net-load ramping, and net-load ramping}.
\citet{Dehghan2015_TPS_GTEP} consider expected unserved energy as a reliability measure. \citet{Moreira2016_TPS_GTEP} incorporates the component failure as a $N-K$ contingency as well as the spatial correlation between nodal demand and supply. \citet{Li2017_TPS_GTEP} capture net load ramping and net load duration uncertainty, considering flexible AC transmission systems (FACTS) devices to provide transmission flexibility. In \citep{Moreira2021_EJOR_GTEP}, the long-term climate trends are modeled as discrete scenarios of the SP while the short-term fluctuation in wind generation is modeled through RO's uncertainty set. A variational autoencoder (VAE) architecture is trained in \citep{Brenner2025_GTEP} to construct a spherical latent space projection of the actual unstructured uncertainty set, allowing to generate new, yet realistic, adversarial scenarios in each iteration of the CCG algorithm.  

DRO formulations have gained traction in recent years.  \citet{Hajebrahimi2020_TSG_GTEP} model the uncertainty of demand, VRE generation, and component failure as part of the DRO ambiguity set using their first moment of historical samples. \citet{Chen2022_TPS_GTEP} use principal component analysis (PCA) to construct a similar ambiguity set considering concentrated solar panels coupled with thermal storage. A bilevel model proposed in \citep{Kang2023_EJOR_GTEP} in which the leader decides interregional (e.g., country) transmission expansion while each region (e.g., country) independently determines the generation expansion plans. Both levels are subject to long- and short-term uncertainties of investment and fuel cost for generators, demand, and VRE generation. Moment-based DRO models long-term uncertainties using the first moment and short-term uncertainties via a budget-box uncertainty set. \citet{Xie2023_GTEP} {propose} a bi-objective model with 1-norm Wasserstein ambiguity set DRO, where the shortfall risk is constrained as a reliability measure.

\begin{table}[!b]
\small 
\caption{Overview of generation and transmission expansion problems under uncertainty in power systems}
\vspace{-0.2cm}
{\scriptsize
\setlength{\tabcolsep}{7pt}
\renewcommand{\arraystretch}{1.1} 
\begin{tabular}{l|llllllll}
\toprule
\multirow{2}{0.08\columnwidth}{Reference} &
\multirow{2}{0.05\columnwidth}{Model}&  \multirow{2}{0.07\columnwidth}{Uncertain parameters}& \multirow{2}{0.05\columnwidth}{Solution, Model}&\multirow{2}{0.12\columnwidth}{Granularity}&\multirow{2}{0.35\columnwidth}{Key aspects}\\ \\
\midrule
\multirow{2}{0.14\columnwidth}{\citet{Dehghan2015_TPS_GTEP}}&\multirow{2}{0.05\columnwidth}{2RO}& \multirow{2}{0.1\columnwidth}{demand, wind gen.}& \multirow{2}{0.05\columnwidth}{BD, {MILP}}  &\multirow{2}{0.12\columnwidth}{1h,~1 yr, 5 rep. hrs |73} &\multirow{2}{0.35\columnwidth}{EUE as reliability measure, tailored BD with primal cuts} \\ \\ 
\midrule
\multirow{2}{0.14\columnwidth}{\citet{Munoz2015_GTEP}}&\multirow{2}{0.05\columnwidth}{2SP}& \multirow{2}{0.1\columnwidth}{demand, VRE gen.}& \multirow{2}{0.05\columnwidth}{PHA, {MILP}}  &\multirow{2}{0.12\columnwidth}{1h,~1 yr, {500 scen.} |240} &\multirow{2}{0.35\columnwidth}{RPS, planning reserve margin} \\ \\ 
\midrule
\multirow{2}{0.14\columnwidth}{\citet{Parkinson2015_GTEP}}&\multirow{2}{0.05\columnwidth}{2SP}& \multirow{2}{0.11\columnwidth}{ave., peak \& flexible dem.}& \multirow{2}{0.05\columnwidth}{EF, LP}  &\multirow{2}{0.12\columnwidth}{1 season,~40 yr |11} &\multirow{2}{0.35\columnwidth}{carbon tax, pumped hydro storage considered, partial load flexibility} \\ \\ 
\midrule
\multirow{3}{0.14\columnwidth}{\citet{Pineda2016_EJOR_GTEP}}&\multirow{3}{0.05\columnwidth}{2SP}& \multirow{3}{0.1\columnwidth}{wind generation}& \multirow{3}{0.05\columnwidth}{EF, MILP}  &\multirow{3}{0.125\columnwidth}{3 yrs each with 10 segments |24} &\multirow{3}{0.35\columnwidth}{RPS, bilevel model, operating reserve, impact of wind generation forecast error} \\ \\ \\
\midrule
\multirow{2}{0.14\columnwidth}{\citet{Moreira2016_TPS_GTEP}}&\multirow{2}{0.05\columnwidth}{2RO}& \multirow{2}{0.12\columnwidth}{demand, VRE gen., other}& \multirow{2}{0.05\columnwidth}{{CCG, MILP}}  &\multirow{2}{0.125\columnwidth}{1h,~1 yr, 1 rep. hr |26} &\multirow{2}{0.35\columnwidth}{RPS, $N-K$ contingency, spinning reserve} \\ \\ 
\midrule
\multirow{2}{0.14\columnwidth}{\citet{Liu2017_TPS_GTEP}}&\multirow{2}{0.05\columnwidth}{mSP}& \multirow{2}{0.11\columnwidth}{cost parameters}& \multirow{2}{0.05\columnwidth}{PHA, {LP}}  &\multirow{2}{0.125\columnwidth}{1h,~1 yr, 30 rep. days |3} &\multirow{2}{0.35\columnwidth}{storage expansion considered} \\ \\ 
\midrule
\multirow{2}{0.14\columnwidth}{\citet{Baringo2017_TPS_GTEP}}&\multirow{2}{0.05\columnwidth}{2RO}& \multirow{2}{0.11\columnwidth}{peak demand fuel cost}& \multirow{2}{0.05\columnwidth}{BD, {MILP}}  &\multirow{2}{0.125\columnwidth}{1h,~1 yr, 48 rep. hr |118} &\multirow{2}{0.35\columnwidth}{} \\ \\ 
\midrule
\multirow{3}{0.14\columnwidth}{\citet{Li2017_TPS_GTEP}}&\multirow{3}{0.05\columnwidth}{2RO}& \multirow{3}{0.13\columnwidth}{dem. ramping \& duration curve}& \multirow{3}{0.05\columnwidth}{CCG, {MILP}}  &\multirow{3}{0.125\columnwidth}{1h,~1 yr|157} &\multirow{3}{0.35\columnwidth}{considered FACTS device, construction period for new assets considered} \\ \\ \\
\midrule
\multirow{2}{0.14\columnwidth}{\citet{Verastegui2019_TPS_GTEP}}&\multirow{2}{0.05\columnwidth}{2RO}& \multirow{2}{0.13\columnwidth}{demand, VRE gen.}& \multirow{2}{0.05\columnwidth}{CCG, {MILP}}  &\multirow{2}{0.125\columnwidth}{1h,~1 yr, 3 rep. days|149} &\multirow{2}{0.35\columnwidth}{expansion of both new assets and existing ones are considered} \\ \\ 
\midrule
\multirow{2}{0.14\columnwidth}{\citet{Hajebrahimi2020_TSG_GTEP}}&\multirow{2}{0.05\columnwidth}{mDRO}& \multirow{2}{0.13\columnwidth}{demand, VRE gen., other}& \multirow{2}{0.05\columnwidth}{EF, MILP}  &\multirow{2}{0.125\columnwidth}{1h,~4 yr, 1 rep. day|118} &\multirow{2}{0.35\columnwidth}{LDR, modeled electrification of transportation and its charging control} \\ \\
\midrule
\multirow{2}{0.14\columnwidth}{\citet{Moreira2021_EJOR_GTEP}}&\multirow{2}{0.05\columnwidth}{2SP, 2RO}& \multirow{2}{0.13\columnwidth}{climate trends, wind gen.}& \multirow{2}{0.05\columnwidth}{CCG, {MILP}}  &\multirow{2}{0.125\columnwidth}{1h,~1 yr, 1 rep. hr|160} &\multirow{2}{0.35\columnwidth}{spinning reserve, $N-K$ contingency} \\ \\ 
\midrule
\multirow{2}{0.14\columnwidth}{\citet{Wogrin2021_GTEP}}&\multirow{2}{0.05\columnwidth}{mSP}& \multirow{2}{0.1\columnwidth}{demand}& \multirow{2}{0.05\columnwidth}{DW, {LP}}  &\multirow{2}{0.12\columnwidth}{1h,~4 stages, 1 rep. hr |5} &\multirow{2}{0.35\columnwidth}{Cournot oligopoly as the follower problem of a sequential Stackelberg game}\\ \\ 
\midrule
\multirow{2}{0.14\columnwidth}{\citet{Bennett2021_GTEP}}&\multirow{2}{0.05\columnwidth}{mSP}& \multirow{2}{0.1\columnwidth}{grid damage}& \multirow{2}{0.05\columnwidth}{EF, LP}  &\multirow{2}{0.13\columnwidth}{1h,~5 stages, 81 scen., 2 rep. days |1} &\multirow{2}{0.35\columnwidth}{storage expansion considered} \\ \\  
\midrule
\multirow{2}{0.14\columnwidth}{\citet{Chen2022_TPS_GTEP}}&\multirow{2}{0.05\columnwidth}{mDRO}& \multirow{2}{0.1\columnwidth}{demand, wind gen.}& \multirow{2}{0.05\columnwidth}{EF, MILP}  &\multirow{2}{0.12\columnwidth}{1h,~5 yrs, 1 rep. day|1354} &\multirow{2}{0.35\columnwidth}{thermal storage, partial load flexibility} \\ \\  
\midrule
\multirow{2}{0.14\columnwidth}{\citet{Kang2023_EJOR_GTEP}}&\multirow{2}{0.05\columnwidth}{mDRO, 2RO}& \multirow{2}{0.125\columnwidth}{demand, VRE gen., other}& \multirow{2}{0.05\columnwidth}{EF, MILP}  &\multirow{2}{0.12\columnwidth}{1yr,~20 yrs |10} &\multirow{2}{0.35\columnwidth}{emissions limit, bilevel model} \\ \\  
\midrule
\multirow{2}{0.14\columnwidth}{\citet{Xie2023_GTEP}}&\multirow{2}{0.05\columnwidth}{dDRO}& \multirow{2}{0.125\columnwidth}{demand, VRE gen.}& \multirow{2}{0.05\columnwidth}{EF, MILP}  &\multirow{2}{0.12\columnwidth}{1h,~1 yr,~1 rep. day |118} &\multirow{2}{0.35\columnwidth}{carbon tax, shortfall risk, storage expansion considered} \\ \\  
\midrule
\multirow{2}{0.14\columnwidth}{\citet{Rintamaki2023_TPS_GTEP}}&\multirow{2}{0.05\columnwidth}{2RO, 2SP}& \multirow{2}{0.125\columnwidth}{demand, VRE gen.}& \multirow{2}{0.05\columnwidth}{EF, MILP}  &\multirow{2}{0.12\columnwidth}{1h,~10 yr,~15 {rep. days} |14} &\multirow{2}{0.35\columnwidth}{emissions limit, storage reservoir considered } \\ \\ 
\midrule
\multirow{2}{0.14\columnwidth}{\citet{Zhang2024_COR_GTEP}}&\multirow{2}{0.05\columnwidth}{mSP}& \multirow{2}{0.125\columnwidth}{demand, VRE gen., other}& \multirow{2}{0.05\columnwidth}{BD, {LP}}  &\multirow{2}{0.12\columnwidth}{0.5h,~3 stages, 729 scen. |5} &\multirow{2}{0.35\columnwidth}{emissions limit, storage expansion considered} \\ \\  
\midrule
\multirow{2}{0.14\columnwidth}{\citet{Zuluaga2024_EPSR_GTEP}}&\multirow{2}{0.05\columnwidth}{2SP}& \multirow{2}{0.125\columnwidth}{demand, VRE gen.}& \multirow{2}{0.05\columnwidth}{{PHA, MILP}}  &\multirow{2}{0.13\columnwidth}{1h, {360 rep. days |8000}} &\multirow{2}{0.35\columnwidth}{RPS, storage expansion considered, transmission loss} \\ \\  
\midrule
\multirow{2}{0.14\columnwidth}{\citet{LeeSun2025_GTEP}}&\multirow{2}{0.05\columnwidth}{2SP}& \multirow{2}{0.125\columnwidth}{demand, VRE gen.}& \multirow{2}{0.08\columnwidth}{level-bundle,{LP}}  &\multirow{2}{0.13\columnwidth}{1h,~52 scen. (weeks) |1493} &\multirow{2}{0.35\columnwidth}{emissions limit, $N-1$ line contingencies, storage expansion, capacity reserve} \\ \\
\midrule
\multirow{2}{0.14\columnwidth}{{\citet{Hole2025_EJOR_GTEP}}}&\multirow{2}{0.05\columnwidth}{{mSP}}& \multirow{2}{0.125\columnwidth}{{wind and hydro gen.}}& \multirow{2}{0.08\columnwidth}{{SDDP, LP}}  &\multirow{2}{0.13\columnwidth}{{time block, 52 stages|11}} &\multirow{2}{0.35\columnwidth}{{hydro reservoir, SDDP using cyclic policy graph}} \\ \\
\midrule
\multirow{2}{0.14\columnwidth}{\citet{Brenner2025_GTEP}}&\multirow{2}{0.05\columnwidth}{2RO}& \multirow{2}{0.125\columnwidth}{demand, VRE gen.}& \multirow{2}{0.05\columnwidth}{{CCG, MILP}}  &\multirow{2}{0.13\columnwidth}{1h,~{1yr, 1 rep. day|6}} &\multirow{2}{0.35\columnwidth}{Use ML to capture non-convex uncertainty sets, storage expansion considered} \\ \\
\midrule
\multirow{2}{0.14\columnwidth}{\citet{Sahin2025_GTEP}}&\multirow{2}{0.05\columnwidth}{2SP}& \multirow{2}{0.125\columnwidth}{component failure}& \multirow{2}{0.05\columnwidth}{EF, MILP}  &\multirow{2}{0.13\columnwidth}{1yr,~28 scen. |663} &\multirow{2}{0.35\columnwidth}{capacity reserve, extreme weather scenarios, investment budget} \\ \\   
\bottomrule
\end{tabular}}
\label{tab:GTEP}
\end{table}

\normalsize

\subsection{Other Energy Infrastructure Expansion Models}
Despite the current interdependence between power and gas systems, and potential coupling with hydrogen in the near future, the literature on interdependent energy expansion planning is sparse. Table~\ref{tab:CEM-PG} shows that most existing studies formulate the problem as either SP or RO. \citet{Shao2017_TPS_PG_GTEP} model the impact of extreme events as a destruction budget used to limit the number of failed transmission lines in the uncertainty set, and restrict the load shedding cost as the resiliency metric to a pre-specified level. \citet{Nunes2018_PG_GTEP} construct discrete scenarios to capture the uncertainty of power demand growth, gas price, and VRE generation. Similarly, \citet{Khaligh2019_PG_GTEP} consider demand growth, wind generation and interest rate as uncertain parameters, and propose an alternating direction method of multipliers (ADMM) approach to preserve the data privacy of power and gas companies. \citet{Riepin2021_PG_GEP} consider the uncertainty of power and gas demand, renewable energy capacity, and fuel and price and CO$_2$ penalty. \citet{Khorramfar2025_EJOR} formulate the joint power-gas problem as DRO with both moment-based and Wasserstein distance ambiguity sets and propose an approximation method that sequentially constructs a feasible solution. 
Finally, \citet{Zhou2024_PH_GTEP} {consider} an interdependent power-hydrogen model with hydrogen demand uncertainty. 

The independent gas infrastructure under uncertainty is considered in a few studies. \citet{Fodstad2016_G_TEP} model the uncertainty of decarbonization policy, CCS technology readiness, and energy efficiency as different scenarios. \citet{Hauser2021_G_TEP} {consider uncertainty in} demand, gas price, and the possibility of gas transit from certain regions. {Extending \citet{Zhang2024_COR_GTEP}}, \citet{Zhang2025_EJOR_PGH_GTEP} consider the coupling between electricity, natural gas, and hydrogen. The short-term uncertainty of demand and VRE supply, and the long-term uncertainty of oil and gas prices, are modeled by a multi-horizon SP. The coupling between energy vectors is realized via modeling electrolyzers, hydrogen storage, fuel cell, steam reforming plants, hydrogen pipelines, and retrofitting natural gas. The resulting problem is solved by refining the Benders stabilization method initially proposed by \citet{Zhang2024_COR_GTEP}.

\begin{table}[!b]
\small 
\caption{Overview of expansion problems under uncertainty for multi-vector energy systems}
\vspace{-0.2cm}
{\scriptsize
\setlength{\tabcolsep}{7pt}
\renewcommand{\arraystretch}{1.1} 
\begin{tabular}{l|lllllllll}
\toprule
\multirow{2}{0.08\columnwidth}{Reference} &
\multirow{2}{0.04\columnwidth}{Vectors Problem}& \multirow{2}{0.04\columnwidth}{Model}&  \multirow{2}{0.07\columnwidth}{Uncertain parameters}& \multirow{2}{0.05\columnwidth}{Solution, Model}&\multirow{2}{0.12\columnwidth}{Granularity}&\multirow{2}{0.3\columnwidth}{Key aspects}\\ \\
\midrule
 \multirow{2}{0.1\columnwidth}{\citet{Fodstad2016_G_TEP}}&\multirow{2}{0.05\columnwidth}{G TEP}&\multirow{2}{0.04\columnwidth}{2SP}& \multirow{2}{0.1\columnwidth}{policy, other}& \multirow{2}{0.05\columnwidth}{EF, QP}  &\multirow{2}{0.15\columnwidth}{1yr, 9 yrs, 8 scen.  |40} &\multirow{2}{0.3\columnwidth}{uncertainty in policy and technology dimensions} \\ \\ 
\midrule
 \multirow{2}{0.1\columnwidth}{\citet{Zhao2017_TPS_PG_GTEP}}&\multirow{2}{0.05\columnwidth}{PG GTEP}&\multirow{2}{0.04\columnwidth}{2SP}& \multirow{2}{0.1\columnwidth}{power/gas demand}& \multirow{2}{0.05\columnwidth}{EF, MILP}  &\multirow{2}{0.15\columnwidth}{1h, 10 rep. hrs, 9 scen.  |118P, 14G} &\multirow{2}{0.3\columnwidth}{linearized gas model} \\ \\ 
 \midrule
 \multirow{2}{0.1\columnwidth}{\citet{Ding2017_TPS_PG_GTEP}}&\multirow{2}{0.05\columnwidth}{PG GTEP}&\multirow{2}{0.04\columnwidth}{mSP}& \multirow{2}{0.1\columnwidth}{power/gas demand}& \multirow{2}{0.05\columnwidth}{EF, MILP}  &\multirow{2}{0.15\columnwidth}{87h, 3 stages, 20 scen.  |118P, 15G} &\multirow{2}{0.3\columnwidth}{linearized gas model with compressor} \\ \\ 
  \midrule
\multirow{2}{0.1\columnwidth}{\citet{He2017_TPS_PG_GTEP}}&\multirow{2}{0.05\columnwidth}{PG GTEP}&\multirow{2}{0.04\columnwidth}{2RO}& \multirow{2}{0.12\columnwidth}{pow. demand, wind gen.}& \multirow{2}{0.05\columnwidth}{CCG, {MILP}}  &\multirow{2}{0.15\columnwidth}{4 blocks, 10 yrs |118P, 20G} &\multirow{2}{0.32\columnwidth}{linearized gas model with compressor, spinning reserve, reliability measures } \\ \\ 
   \midrule
 \multirow{2}{0.1\columnwidth}{\citet{Shao2017_TPS_PG_GTEP}}&\multirow{2}{0.05\columnwidth}{PG GTEP}&\multirow{2}{0.04\columnwidth}{2RO}& \multirow{2}{0.11\columnwidth}{num. failed component}& \multirow{2}{0.05\columnwidth}{CCG, {MILP}}  &\multirow{2}{0.15\columnwidth}{1yr, 6 yrs |24P, 17G} &\multirow{2}{0.3\columnwidth}{planning reserve, load shedding as resilience metric } \\ \\ 
    \midrule
 \multirow{2}{0.1\columnwidth}{\citet{Nunes2018_PG_GTEP}}&\multirow{2}{0.05\columnwidth}{PG GTEP}&\multirow{2}{0.04\columnwidth}{2SP}& \multirow{2}{0.12\columnwidth}{power demand, other}& \multirow{2}{0.05\columnwidth}{EF, MILP}  &\multirow{2}{0.15\columnwidth}{1h, 24 rep. hrs, 9 scen. |{11P, 10G}} &\multirow{2}{0.3\columnwidth}{carbon tax, RPS } \\ \\ 
     \midrule
 \multirow{2}{0.14\columnwidth}{\citet{Khaligh2019_PG_GTEP}}&\multirow{2}{0.05\columnwidth}{PG GTEP}&\multirow{2}{0.04\columnwidth}{2SP}& \multirow{2}{0.12\columnwidth}{demand growth, other}& \multirow{2}{0.05\columnwidth}{ADMM, {MILP}}  &\multirow{2}{0.15\columnwidth}{1d, {15} yrs, 48 scen. |17P, 16G} &\multirow{2}{0.32\columnwidth}{partial demand flexiblity, Weymouth equation with compressor } \\ \\ 
      \midrule
 \multirow{2}{0.14\columnwidth}{\citet{Riepin2021_PG_GEP}}&\multirow{2}{0.05\columnwidth}{PG GEP}&\multirow{2}{0.04\columnwidth}{2SP}& \multirow{2}{0.12\columnwidth}{pow/gas demand, other}& \multirow{2}{0.05\columnwidth}{EF, LP}  &\multirow{2}{0.15\columnwidth}{1h, 350 hourly scen. |16PG} &\multirow{2}{0.3\columnwidth}{node-specific constraints such as allowance for nuclear expansion} \\ \\ 
\midrule
\multirow{2}{0.1\columnwidth}{\citet{Yamchi2021_PG_GTEP}}&\multirow{2}{0.05\columnwidth}{PG GTEP}&\multirow{2}{0.04\columnwidth}{2SP}& \multirow{2}{0.11\columnwidth}{pow. load, solar gen.}& \multirow{2}{0.05\columnwidth}{EF, MILP}  &\multirow{2}{0.16\columnwidth}{20yrs into 4 block, 12 scen. |24P, 20G} &\multirow{2}{0.3\columnwidth}{linearized gas model, carbon tax, $N-1$ contingency
} \\ \\ 
\midrule
\multirow{2}{0.1\columnwidth}{\citet{Hauser2021_G_TEP}}&\multirow{2}{0.05\columnwidth}{G TEP}&\multirow{2}{0.04\columnwidth}{2SP}& \multirow{2}{0.11\columnwidth}{demand, other}& \multirow{2}{0.05\columnwidth}{EF, LP}  &\multirow{2}{0.15\columnwidth}{1day, 3 periods, 12 scen. |47} &\multirow{2}{0.3\columnwidth}{gas storage considered} \\ \\ 
\midrule
\multirow{2}{0.1\columnwidth}{\citet{Riepin2022_G_TEP}}&\multirow{2}{0.05\columnwidth}{G TEP}&\multirow{2}{0.04\columnwidth}{2RO}& \multirow{2}{0.11\columnwidth}{demand, gas supply}& \multirow{2}{0.05\columnwidth}{CCG, {MILP}}  &\multirow{2}{0.15\columnwidth}{1 month, 1 yr |37} &\multirow{2}{0.3\columnwidth}{gas storage considered} \\ \\ 
\midrule
\multirow{2}{0.1\columnwidth}{\citet{Zhou2024_PH_GTEP}}&\multirow{2}{0.05\columnwidth}{PH GTEP}&\multirow{2}{0.04\columnwidth}{2RO}& \multirow{2}{0.11\columnwidth}{demand, gas supply}& \multirow{2}{0.05\columnwidth}{CCG, {MILP}}  &\multirow{2}{0.15\columnwidth}{1h, 4 periods, 4 rep. days |13} &\multirow{2}{0.3\columnwidth}{hydrogen storage considered} \\ \\ 
\midrule
\multirow{2}{0.1\columnwidth}{{\citet{Goke2024_EJOR_GTEP}}}&\multirow{2}{0.05\columnwidth}{{PGH GTEP}}&\multirow{2}{0.04\columnwidth}{{2SP}}& \multirow{2}{0.11\columnwidth}{{demand, VRE gen.}}& \multirow{2}{0.05\columnwidth}{{BD, MILP}}  &\multirow{2}{0.15\columnwidth}{{1h, 2yrs, 16 scen.|4PGH}} &\multirow{2}{0.3\columnwidth}{{emissions limit, pumped hydro and hydrogen storage considered}} \\ \\ 
\midrule
{{\citet{Kayacik2025_PH}}}&\multirow{2}{0.05\columnwidth}{{H GEP}}&\multirow{2}{0.04\columnwidth}{{mDRO}}& \multirow{2}{0.11\columnwidth}{{demand}}& \multirow{2}{0.05\columnwidth}{{CCG, MILP}}  &\multirow{2}{0.15\columnwidth}{{5yrs, 20yrs|19}} &\multirow{2}{0.3\columnwidth}{{speed-up techniques}} \\ \\ 
\midrule
\multirow{2}{0.1\columnwidth}{\citet{Khorramfar2025_EJOR}}&\multirow{2}{0.05\columnwidth}{PG GTEP}&\multirow{2}{0.04\columnwidth}{mDRO dDRO}& \multirow{2}{0.11\columnwidth}{demand, VRE gen.}& \multirow{2}{0.05\columnwidth}{EF, MILP}  &\multirow{2}{0.15\columnwidth}{1h, 1 yr, 15 rep. days |6} &\multirow{2}{0.3\columnwidth}{CVaR, RPS, battery storage expansion considered} \\ \\ 
\midrule
\multirow{2}{0.1\columnwidth}{\citet{Zhang2025_EJOR_PGH_GTEP}}&\multirow{2}{0.05\columnwidth}{PGH GTEP}&\multirow{2}{0.04\columnwidth}{mSP}& \multirow{2}{0.11\columnwidth}{demand, other}& \multirow{2}{0.05\columnwidth}{BD, {MILP}}  &\multirow{2}{0.16\columnwidth}{1h, 4 rep. periods,  {7} stages |27} &\multirow{2}{0.3\columnwidth}{expansion of pumped hydro and battery storage, carbon tax} \\ \\ 
\bottomrule
\end{tabular}}

\vspace{0.1cm}
{\scriptsize \hspace{4cm} In the second column, \textit{P, G}, and \textit{H} refers to electric power, gas, and hydrogen systems}
\label{tab:CEM-PG}
\end{table}

\normalsize

\subsection{Open-source Capacity Expansion Modeling Tools}
Open-source CEM tools provide a practical lens through which many of the methodological choices discussed in this section are implemented. Examining these tools reveals how theoretical modeling advances translate into planning tools, and where important gaps persist. To identify representative tools, we queried the \textit{Open Energy 
Transition Tool Tracker} \cite{openmod_tracker_dashboard}, and applied several screening criteria: i) publicly available documentation; ii) active maintenance as of May 2025; iii) engagement on GitHub (answering questions, debugging); iv) minimum package download within the past month; and v) no reliance on proprietary software. We additionally include all tools explicitly tagged as ``capacity-expansion'' in the database, regardless of engagement metrics. We omit the tools that meet these criteria but lack sufficient documentation or fall outside of the paper's scope. Table~\ref{tab:tools} summarizes the identified tools. 
All tools reviewed in this section are open source, primarily used in research institutions. A few tools such as PyPSA and ReEDS are employed in national studies by \citet{cer_ef2023_appendix3, gagnon2024standard}. {In practice, utilities and agencies also rely on commercial platforms such as PLEXOS and EnCompass, which link long-term capacity expansion with production cost and other operational analyses \citep{EPRI2024_Tools}. For example, AES Indiana uses EnCompass for capacity expansion with hourly production-cost modeling in its 2025 integrated resource plan where it combines deterministic scenarios with stochastic evaluation of candidate portfolios \citep{AESIdiana2025_tools}. Here, however, Table~\ref{tab:tools} focuses on open-source tools due to licensing restrictions and limited public documentation that prevent a consistent and reproducible comparison. }


\begin{table}[!b]
\small
\caption{Tooling overview for capacity and transmission expansion modeling under uncertainty}
\label{tab:tools}

{\scriptsize
\setlength{\tabcolsep}{6pt}
\renewcommand{\arraystretch}{1.12}
\renewcommand{\tabularxcolumn}[1]{m{#1}}

\begin{tabularx}{\columnwidth}{m{0.09\columnwidth}|y{0.12\columnwidth}|C{0.09\columnwidth}C{0.06\columnwidth}C{0.1\columnwidth}C{0.1\columnwidth}Y}
\toprule
Tool & Reference & Language & Model & {Temporal Aggregation} & Multistage & Key aspects \\
\midrule
 PyPSA & \cite{PyPSA} & Python & 2SP & \xmark & PF/Myopic &
Water-inflow, risk measures, UC, reserves. \\
\midrule
Temoa & \cite{Temoa} & Python & 2SP + MCS & \xmark & RH/PF/ Myopic &
Reliability, reserves. \\
\midrule
Tulipa & \cite{tulipa_energy_model}& Julia & 2SP & \cmark & RH/PF/ Myopic &
Water-inflow, reliability, UC. \\
\midrule
FINE & \cite{FINE} & Python & 2SP & \cmark & RH/PF/ Myopic &
Water-inflow, UC. \\
\midrule
flixOpt & \cite{flixopt} & Python & 2SP & \cmark & RH/PF/ Myopic &
Multisector; risk measures, UC.\\
\midrule
SpineOpt & \cite{SpineOpt} & Julia & mSP & \cmark & RH/PF/ Myopic &
BD; water-inflow, reliability, UC, reserves. \\
\midrule
ReEDS & \cite{REEDS}& \parbox[t]{\linewidth}{\centering Julia +\\GAMS+Py} & Det + MCS& \cmark & PF/Myopic &
MGA for finding near-optimal solutions; reliability, reserves. \\
\midrule
GridPath & \cite{gridpath2022}& Python & Det + MCS & \xmark & RH/PF/ Myopic &
Resource-adequacy toolkit; reliability, UC, reserves. \\
\midrule
IESopt & \cite{iesopt} & Julia & Det + MCS & \xmark & PF/Myopic &
Water-inflow. \\
\midrule
GenX & \cite{genx}& Julia & Det + SA & \cmark & PF/Myopic &
Method of Morris SA; water- inflow, reliability, UC, reserves. \\
\midrule
Macro & \cite{macdonald2025macroenergy}& Julia & Det & \cmark & RH/PF/ Myopic &
Multisector; BD; water-inflow, reliability, UC, reserves. \\
\midrule
AnyMOD & \cite{AnyMod}& Julia & {2SP} & \xmark & PF &
{BD, water-inflow, reliability.} \\
\midrule
Calliope & \cite{pfenninger2018Calliope} & Python & Det & \cmark & PF & Water-inflow, reserves. \\
\midrule
REMix & \cite{wetzel2024REMix} & GAMS+Py & Det & \xmark & PF &
Multi-commodity; water-inflow, UC. \\
\midrule
PREP-SHOT & \cite{liu2023} & Python & Det  & \xmark & PF &
Hydropower-centric; water-inflow. \\
\midrule
OSeMOSYS & \cite{howells2011OSeMOSYS} & GMPL & Det & \cmark & PF &
Water-inflow, reliability measures; reserves. \\
\midrule
{HOPE} & {\cite{wang2025hope}} & {Julia} & {Det} & {\cmark} & {PF} & {Reliability,
reserves; LLM interface.}\\
\midrule
ZEN-garden & 
\cite{ZENgarden2025} & Python & Det & \cmark & PF/Myopic &
\\
\midrule
URBS & \cite{urbs} & Python & Det & \xmark & PF & \\
\midrule
solph & \cite{krien2020oemofsolphA} & Python & Det & \cmark & PF/Myopic &
 \\ 
\bottomrule
\end{tabularx}}
\end{table}


\vspace{0.15cm}
\noindent \textbf{Modeling considerations.}
Despite differences in terminology, the reviewed tools share a common structural representation of energy systems. Most frameworks model the system as a network of generation sources, demand sinks, energy transport (e.g., transmission lines), and storage technologies. For sensitivity analysis, these tools typically require inputs such as investment costs, fixed and variable costs, emissions limits, load, and renewable outputs. Several tools extend this baseline representation with additional representations, such as water inflow for hydro reservoir storage, reliability metrics, UC constraints, reserve considerations, and risk measures, as shown in column ``Key aspects'' of Table~\ref{tab:tools}. 

Temporal resolution is typically user-specified and can easily be adjusted. {Many tools also support temporal aggregation, in which hours or days with similar demand and renewable-generation profiles are clustered into a smaller set of weighted representative periods that capture seasonal variability.} All tools support a perfect foresight (PF) mode where the future system condition is fully known. Many tools also allow myopic planning (decisions are made without full foresight), or rolling-horizon (RH) with limited look-ahead over a moving time window. {Spatial resolution is often configurable through the input network, but choice of resolution is not completely free. Feasible resolution depends on the tool’s supported network formulation, data requirements, available datasets, and scalability.} Some platforms adopt predefined spatial structure; for example, ReEDS uses mixed-resolution balancing area and county representation and GenX employs a zonal model. While most tools focus on electricity systems planning, a subset supports multi-vector energy modeling. Examples include flixOpt, which co-optimizes electricity and heating systems, and REMix, which includes hydrogen, synthetic fuels, and gas network through modular extensions.

\vspace{0.15cm}
\noindent \textbf{Uncertainty handling.} Across the open-source CEM tools reviewed here, uncertainty is most commonly represented through scenario enumeration. Several tools support sensitivity analysis across alternative input assumptions by solving multiple deterministic model instances. 
Some frameworks complement this approach with sampling-based techniques, such as Monte Carlo. Beyond sensitivity analyses, a smaller subset of tools implement scenario-coupled stochastic programming, most often in a two-stage form in which investment decisions are shared across scenarios while operations vary by scenario. Only limited support exists for more advanced stochastic structures. Among tools, SpineOpt provides one example of a multi-stage SP using directed acyclic graphs that allows branching and convergence of scenarios over time. 

\vspace{0.15cm}
\noindent \textbf{Problem formulations and solvers.} 
Most reviewed tools implement LP as a baseline and allow MILP extension when integer variables are required. Many tools are solver-agnostic and use packages such as JuMP, Linopy, or Pyomo to interface with external solvers. Some tools also provide additional algorithmic support to improve scalability. For example, Macro includes decomposition-based solution techniques for large-scale problems. Nevertheless, support for a particular formulation does not guarantee tractability, as solving realistic instances depends heavily on solver performance and model characteristics.

\section{Challenges and Research Gaps}\label{sec:gaps}
Despite the substantial progress in modeling and data-driven approaches, energy infrastructure planning under uncertainty remains fraught with methodological and practical challenges. Increasing system complexity has expanded both the scale of optimization models and the scope of uncertainties they should represent. As a result, many models simply approximate uncertainty in an ad hoc manner, and face computational barriers when scaled to realistic systems.
This section identifies these challenges and outlines research gaps where the decision value of added fidelity is theoretically motivated but empirically unverified, where computational barriers prevent exploring the fidelity-decisions relationship, and where methodological innovation may thereby be needed. The discussion is organized along four themes--modeling fidelity, uncertainty characterization, solution approaches, and validation and decision impact---each detailing where current methods fall short and outlining potential directions for future research. The goal is to chart a path for decision-making models that can support the decision making in different contexts to achieve sustainable energy infrastructure planning.  
\subsection{Modeling Fidelity}\label{ssec:fidelity}
In energy systems, the modeling choices play a critical role in ensuring that the resulting outcomes are realistic and align with the planning goal. The scope and assumptions of a model dictate which technical constraints must be represented to guarantee the feasibility and consistency with the objective function. Although the literature captures many key technical aspects, several gaps remain that warrant further investigation. 

\vspace{0.15cm}

\noindent \textbf{Integrated Power-Gas-Hydrogen Planning.} 
Energy vectors are increasingly interconnected  \citep{Klatzer2022_survey, OrdoudisEtal2019_PG_EJOR}. Power systems rely on gas-fired generators to balance renewables, and gas systems rely on power systems to operate components that maintain gas flow and pressure; hydrogen is also gaining traction as both an energy carrier and storage medium, tying its operations to both power and gas systems  \citep{He2021_H2Power}. Added to this are the ambitious goals to electrify heating and transportation, which create an even tighter coupling, thus making it imperative to navigate the sustainable transition landscape with an integrated approach. From an operations research perspective, integrated power-gas-hydrogen systems can require fundamentally new modeling challenges. Hydrogen enables bidirectional coupling between energy carriers through technologies, such as electrolyzers and fuel cells, that result in further spatio-temporal interdependencies. 
Although some studies incorporate these interactions (Table~\ref{tab:CEM-PG}), further work is imperative to investigate how multi-vector planning alters expansion pathways under weather-induced, technological, and socio-economic uncertainties. With growing adoption of VRE and electrification of end-use, integrated models can capture cross-sectoral synergies, such as retrofitting gas networks, hydrogen as long-duration storage, power-to-gas technologies, and hydrogen infrastructure.

\vspace{0.15cm}
\noindent \textbf{Coordinated Transmission and Generation Planning in Power Systems.} Much of the literature still treats generation and transmission in isolation. Yet, from a central planner's perspective, coordinated GTEPs are essential as they explicitly model the interdependencies between electricity generation points and power delivery to consumers \citep{LiEtal2022_EJOR}. This is especially important in highly decarbonized systems where transmission expansion is often the bottleneck for renewable integration \citep{Brown2021_Joule}. 
Conversely, transmission expansion without commensurate demand or generation investment can result in costly and underutilized assets. Coordinated models enable planners to identify more cost-effective transition pathways and improve adaptability toward uncertainty and disruptions by better responding to evolving demand and supply patterns.
Future research should quantify and compare coordinated and isolated models---especially in terms of system cost,  reliability, and robustness to extreme events--- to determine when GTEP provides significant decision value relative to its added value. 

\vspace{0.15cm}
\noindent \textbf{Coupling with Other Infrastructures.}
Energy systems are deeply interwoven with other infrastructures, creating layers of interdependence. Water, for instance, depends on electricity for extraction, treatment and distribution while thermal generators and data centers rely on water for cooling \citep{Bauer2014_DOE_Water}. {These interdependencies also extend to environmental systems. \citet{Xu2025_hydro} integrates a CEM with hydropower operations and sediment transport, and show how how infrastructure decisions can create consequential trade-offs between decarbonization costs adn downstream ecosystem services.} Transportation exhibits a similar bidirectional relationship: all modes of transportation rely on energy, and in return, transportation electrification is a major part of decarbonization strategies \citep{Khargonekar2024}. Despite these links, only \cite{Hajebrahimi2020_TSG_GTEP} consider power-transportation coupling. Its interependencies also extend beyond physical networks to finance, communication, and digital infrastructure, all of which rely on electricity. 
These connections emphasize the need for a more comprehensive approach that incorporates and reflects the structural relationships across infrastructures. 

\vspace{0.15cm}
\noindent \textbf{Spatio-temporal Resolution.}
Many existing models adopt coarse spatial or temporal granularity by aggregating locations (e.g., nodes in a county or state) or relying on representative periods for primarily two reasons: i) computational tractability; ii) to preserve the decision boundary (e.g., each administration unit knows its planning outcome). 
However, these simplifications can obscure critical planning dynamics. Coarse spatial resolution can mar the complementarity effect between solar generation, wind generation and transmission congestion, leading to biased cost and reliability estimates \citep{Serpe2025_NREL, Qiu2024_CRS}.
Under uncertainty, insufficient resolution can also distort risk assessments. For instance, aggregated temporal profiles smooth peak loads and mask extreme events, thus concealing the tail risk \citep{Qiu2024_CRS}. High-resolution supply and demand data are therefore important for designing robust and cost-effective energy systems. Beyond parameters, fine-grained data facilitates just transition policies, allowing planners to better detect, quantify, and mitigate distributional disparities. Future work should aim to integrate high-resolution data while preserving the tractability, for example through advances in decomposition and parallel computing (e.g., GPU-accelerated optimization).

\vspace{0.15cm}
\noindent \textbf{Market Structures and Strategic Interactions.}
The institutional setting in which infrastructure is planned strongly influences investment and operational outcomes. In vertically integrated power systems, a central planner coordinates expansion across the supply chain. In contrast, deregulated markets decentralize transmission and generation decisions, which potentially creates misaligned incentives and strategic interactions. {Therefore, uncertainty in this context extends beyond input parameters and includes market and regulatory design, degree of coordination among jurisdictions, and any decision for strategic participants. These institutional and behavioral uncertainties can result in substantially different system-level and distributional effects. \citet{Kucuksayacigil2025} examine alternative levels of market and policy coordination in the western United States, and find that full coordination mildly reduces the aggregate system cost but can greatly impact state-level cost. Coordination also changes the location and composition of investment. \citet{Gonzalez2021} compare transmission and storage expansion under a social planner and a merchant investor, and show that while aggregate social welfare is relatively small, the generation mix can change substantially.}

{Related concerns arise in gas and emerging hydrogen markets. Gas infrastructure is characterized by concentrated production and transmission ownership, long-term contracts, and strategic interactions among upstream producers, pipeline operators, and terminals \citep{Abada2017_gas_market, Zahur2023_gas_market}. Relevant uncertainties include fuel and transport prices, contract terms, pipeline access, market power, and the response of competing suppliers to new capacity. }
Hydrogen infrastructure planning is in its nascent stages, and uncertainties in technology development and market formation can lead to strategic behavior in siting production, transportation, storage, and demand clusters.

{Bilevel, equilibrium, and game-theoretical formulations can represent strategic interactions by allowing one actor's investment or policy decision to anticipate the responses of other participants \citep{Pineda2016_EJOR_GTEP, Wogrin2021_GTEP, Kang2023_EJOR_GTEP, Garcia2025_EJOR_GEP}. Nevertheless, few uncertainty-aware CEMs combine strategic behavior with uncertain market rules, policy coordination, prices, and participant responses. Future research should therefore evaluate alternative institutional structures using not only aggregate cost and welfare but also investment outcomes and allocation of benefits and risks among stakeholders.}

\vspace{0.15cm}
\noindent \textbf{Policy and Regulatory Mandates.} 
Policy and regulation strongly shape planning by setting priorities, directing investments, and ultimately impacting both system operations and long-term visions. Instruments such as RPS, methane leakage standards, and the recent US Inflation Reduction Act \citep{Bistline2023_IRA} create strong incentives for low-carbon technologies. Nevertheless, these mandates are often stylized in CEMs, imposed through simplified constraints (e.g., RPS) or exogenous parameter adjustments (e.g., carbon tax). This abstraction limits the ability of models to capture broader system impacts. For example, policies that encourage renewable integration may inadvertently impact system resiliency due to insufficient firm power or storage facilities. Moreover, the distributional effects are rarely modeled, despite policies such as net-metering \citep{Sunar2021_Net_Metering} and community solar \citep{O2024_NE_comm_solar} having important equity and affordability implications across regions and consumer groups. 

Future research should develop uncertainty-aware policy representation that accounts for evolving policy trajectories (e.g., subsidy regimes). Apart from costs and investment outcomes, models should also evaluate the distributional and resilience implications of mandates and explore tradeoffs between overlapping policies, such as local incentives for heating electrification and national decarbonization targets. 


\vspace{0.15cm}
\noindent \textbf{Construction and Interconnection Delays.} 
CEMs typically assume that all investments are built and commissioned on schedule. In practice, permitting bottlenecks, financing, supply chain disruption, or social opposition can substantially delay projects. In power systems, transmission expansion often faces right-of-way and siting barriers, while many generation projects encounter long interconnection queues, 
particularly in the US where most proposed projects eventually withdraw \citep{Gorman2025_Joule}. 
Despite their significance, such delays are rarely modeled endogenously in CEMs. Recently, \citet{Glista2024_Interconnection_Queue} proposed a deterministic GEP in which interconnection delays probabilistically impact the success rate of new assets, limiting additions of each storage and generator type to a certain predicted number. 
Similar challenges arise in gas and hydrogen infrastructure, where pipeline projects face permitting challenges, cross-border (or intra-state in the case of US) negotiations, and public resistance. Therefore, integrating the uncertainty of project realization into infrastructure planning remains a research frontier. 

\vspace{0.15cm}
\noindent \textbf{Energy Storage.} 
Storage is a key enabler of low-carbon energy systems because it improves reliability and resilience under growing uncertainty. By decoupling production from consumption, storage helps mitigate the intermittency of VRE, provide grid services (peak shaving, grid stability, and resiliency), reduce investment in new assets, and dampen price volatility \citep{Levin2023_survey, Sioshansi2021_survey_storage}. Technologies range from batteries and pumped hydro to thermal storage and power-to-hydrogen systems. Despite its indispensability, storage is underrepresented in the literature, with fewer than a third of the studies surveyed including it. Important gaps remain in modeling uncertainty of storage-related parameters (cost, round-trip efficiency), assessing the system impact of long-duration storage (i.e., technologies capable of discharging for more than 10 hours), and incorporating thermal and hydrogen storage in integrated multi-vector models. These technologies can significantly influence system costs, reliability, and performance under future scenarios \citep{Dowling2020_Joule}.

Gas and hydrogen storage are even less studied.  Yet, gas storage has long provided flexibility for demand fluctuations through underground reservoirs and pipeline linepack. The ongoing transition toward decarbonization introduces new complexities including the optimal use of existing storage assets in light of declining gas and increasing hydrogen consumption. 
Hydrogen storage introduces further uncertainty due to evolving technology, cost, and location feasibility ( salt cavern, tanks, or chemical carriers).
Therefore, more detailed treatment in gas and hydrogen systems is needed to account for alternative use cases for current facilities and the rollout of new technologies across both vectors. 

\vspace{0.15cm}
\noindent \textbf{Energy Transport.} Modeling the transmission and transportation of energy is a key aspect of CEMs \citep{Ruiz2015_EJOR_TEP, AndreEtal2013_EJOR}. In power systems, transmission constraints are often approximated by lossless transportation or DC models because solving full AC power flow yields computationally challenging mixed-integer nonlinear programs. Some studies approximate transmission losses by simplified formulations such as proportional loss factor and linear approximation \citep{Guevara2020_AE_GEP, Irawan2023_EJOR_GEP, Zuluaga2024_EPSR_GTEP}. 
However, oversimplifying representations may or may not distort planning outcomes.
Depending on the problem, DC models may underestimate investment costs and produce solutions that violate AC constraints \citep{bent2014Transmissiona}, or provide a good approximation \citep{KhanpourMolzah2026}. 
While \citet{Neumann2022_AE} compare alternative power flow representations in deterministic settings, a systematic assessment of these aspects under uncertainty is still missing. Understanding how transmission network modeling influences planning outcomes and computational tractability remains an open research gap. 

Gas and hydrogen networks present different challenges. Although their flows follow nonlinear compressible fluid dynamics, planning models rely on linearized approximations \citep{Ding2017_TPS_PG_GTEP, Riepin2022_G_TEP, Zhou2024_PH_GTEP, Zhang2025_EJOR_PGH_GTEP}. Additional characteristics, such as slower gas flow, linepack storage, and alternative hydrogen transport modes, are rarely modeled. {Hydrogen transportation also introduces trade-offs among discrete and continuous transport decisions}. 
There are unique dimensions for gas and hydrogen as well. Gas transmission is slower and offers storage capacity within pipelines (linepack). Besides pipelines, hydrogen can be transported by truck and ship, each differing in cost, efficiency and scale. None of these aspects has been considered in the literature, but these modeling choices can significantly affect tractability and solution quality.  
     
\vspace{0.15cm}
\noindent \textbf{Operation of Thermal Power Generators.} Conventional power plants such as gas and nuclear units are subject to technical constraints including UC, ramping, and minimum up/down time. While these are often represented in  GEPs, they are mainly simplified or omitted in TEPs and GTEPs. 
However, such constraints can significantly impact systems operations and investment outcomes \citep{Schwele2020_UC_in_GEP}, especially under the increasing importance of dispatchable generation in balancing variable renewables. More systematic assessment of unit-specific constraints and their associated costs are needed.

\vspace{0.15cm}
\noindent \textbf{Reserve Requirements in Power Systems.} Reserves provide essential flexibility and reliability under uncertainty, but only a handful of studies explicitly model them. Existing work examines operating reserves in a deterministic setting \citep{Van2016_TPS}, but their role under uncertainty remains unexplored. Furthermore, reserve types differ in relevance across planning horizons and decision context; spinning reserves are crucial for system adequacy, whereas non-spinning or frequency-response reserves are more relevant in shorter time slices and distribution level. Extending reserve modeling into stochastic CEMs is an important future direction.  

\vspace{0.15cm}
\noindent \textbf{Demand Side Measures in Power Systems.} By allowing customers to shift or curtail their consumption in response to market signals, demand response programs can shape both operational and investment decisions \citep{GarttnerEtal2018_EJOR}. As a flexible resource, it can reduce peak demand, defer network upgrades, and mitigate the need for dispatchable power generation \citep{Motta2024_survey_DRP, Chen2022_TPS_GEP}. Nevertheless, most studies treat demand as inelastic, neglecting how incentive-driven consumption may evolve under large-scale VRE integration, decarbonization targets, and electrification goals. 
Demand flexibility also introduces uncertainty across heterogeneous consumer groups and regions, making it well-suited for representation through SP, RO, or DRO.  Furthermore, participation levels depend on policy, technology, and behavioral factors that evolve endogenously in response to infrastructure investment and market design \citep{Giannelos2025_TEP}. Therefore, future studies should model demand flexibility more explicitly, including its reliability and equity implications, and its potential dependence on infrastructure.






\subsection{Uncertainty Characterization} \label{ssec:uncertainty_char}
\noindent \textbf{Discrete Scenarios vs. Uncertainty Sets.}
Stochastic CEMs are often categorized by whether they adopt a scenario-based or a robust formulation. Scenario-based approaches minimize expected or risk-adjusted costs over a finite set of scenarios, representing possible realizations of uncertain planning parameters. Robust formulations instead minimize costs under worst-case realizations within the uncertainty set. Although these approaches reflect different paradigms, the choice between them is often driven by practical considerations, most notably the availability and reliability of data describing future conditions. Over long planning horizons, or when historical data are limited, scenario-based formulations may inadequately capture tail risk. In such cases, robust formulations offer a conservative yet tractable way for hedging against deep uncertainty. 
To balance robustness and conservativeness, some studies adopt hybrid approaches. A common approach is to ``robustify'' scenario-based (or deterministic) models by adding extreme scenarios \cite{li2022representative}. While computationally tractable, this approach may limit robustness to a set of predefined scenarios and may fail to capture risks beyond historical observations.  

\vspace{0.15cm}
\noindent \textbf{Endogenous Uncertainty.}
{Most CEMs treat uncertainty as exogenous where the possible realization and their probabilities are independent of the model's investment decisions. Endogenous, or decision-dependent, uncertainty instead arises when a decision affects either the uncertainty process itself or the information available about it \citep{Apap2017_DD_uncertainty}. The form is decision-dependent distributional uncertainty in which an investment or policy decision changes the probability, support, or magnitude of an uncertain parameter. For example, installed generation capacity can affect the distribution of future energy prices \citep{Zhan2016_TPS_GEP}. The second form is decision-dependent information revelation where the underlying outcome does not necessarily change but an investment determines the timing or quality of its observability. \citet{Giannelos2025_TEP}, for instance, model network investment in smart technologies that reveal previously uncertain information about demand response participation.}

{Despite its relevance, endogenous uncertainty remains uncommon in CEMs because specifying a credible causal relationship between investment and uncertainty requires additional data and assumptions, and often produces computational hurdles.} However, modeling endogenous uncertainty would better capture feedback loops between infrastructure decisions and changing system conditions including the co-evolution of supply and demand in deeply decarbonized systems; {the investment changes the uncertain environment, which changes the expected or worst-case operational outcomes, which in return impacts the quality and value of the investment itself.} 

\begin{figure}
    \centering
    \includegraphics[width=0.8\linewidth]{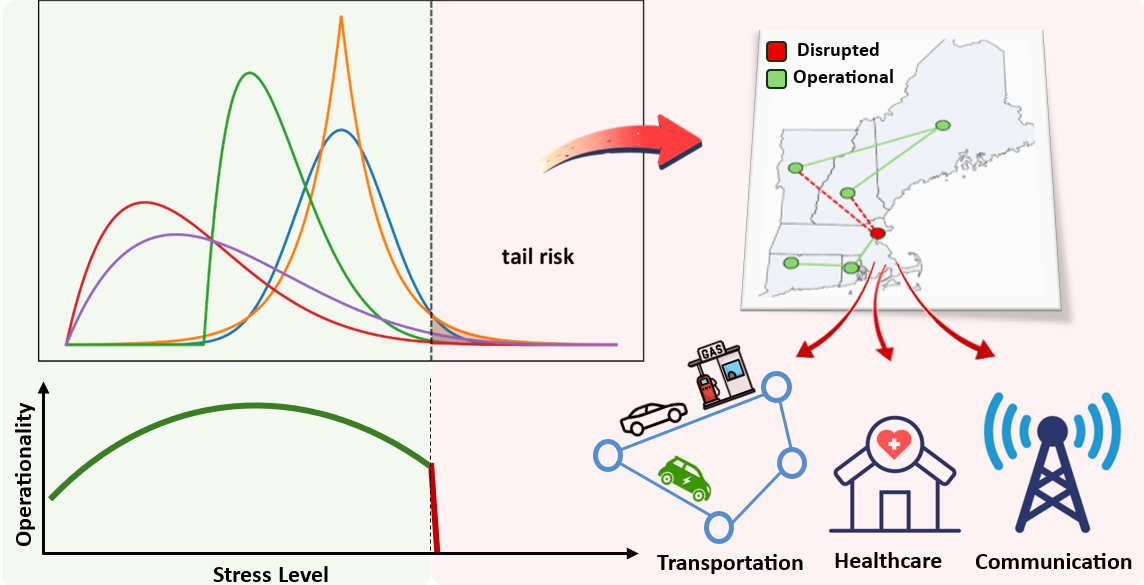}
    \caption{{Generalized propagation of} tail event-induced disruption {across energy systems and dependent services.} to other systems. In the left part, in pale green, the system functions even though the distribution of the stochastic parameter is uncertain. In the right pale red area, the system quickly fails in response to a tail risk, disrupting components of energy systems (top, right), which in turn renders critical infrastructure and services such as transportation, healthcare, and communication dysfunctional.}
    \label{fig:tail-risk}
    \vspace{-1em}
\end{figure}

\vspace{0.15cm}
\noindent \textbf{Risk, Resiliency, and Contingency Analysis.}
Another underexplored dimension in CEMs is tail risk, i.e., low-probability but high-impact events {whose consequences can be highly asymmetric.}
However, these events are inherently data-poor, thus making it difficult for RO and DRO models to reliably capture the frequency and severity of extreme events. {Within the current optimization paradigms, risk measures and constraints such as CVaR and chance constraints can make risk tolerance explicit.} Although some studies incorporate extreme events  \citep{Bennett2021_GTEP}, only a small share explicitly employ such tools. 
The critical aspect of tail risk lies not in the initial extreme event per se, but in the chain of events it can trigger. Disruptions such as as wind-induced line failure can cause cascading impact throughout the energy systems, turning a localized disruption into systemic crises \citep{Xu2025_joule_survey, Sturmer2024_NE, Quiggin2021_Chatham_House}. {Fig.~\ref{fig:tail-risk} presents this general propagation mechanism where a failure in one part of an energy network can disrupt other energy components and subsequently vital services. The Texas power crisis during Winter Storm Uri exemplifies this mechanism, as the interaction between gas shortages and constrained power generation contributed to severe power outages and disruption to heating, water, and medical services \citep{Busby2021_cascading}. Evidence from climate-risk assessments similarly shows that ignoring tail risk can understate losses and misrepresent resilience across interdependent systems \citep{Quiggin2021_Chatham_House}}.

{Tail events can be incorporated through stress scenarios, importance sampling, risk measures, and enhanced contingency analysis. } However, current studies may only enforce $N-1$ and $N-2$ contingencies due to computational limits \citep{LeeSun2025_GTEP}. Extending planning models to consider larger contingency sets and cascading effects while maintaining tractability remains an open research challenge.

\subsection{Solution Approach} \label{ssec:sol_approach}
The increasing complexity of stochastic CEMs has warranted tailored solution mechanisms as they become large-scale convex or non-convex problems that are difficult to solve \citep{Kaya2025_EJOR_survey}. {Models with coarse granularity (less than 10 nodes and few representative days) can often be solved in extensive form as MILP or MISOCP (Table~\ref{tab:GEP}). As spatio-temporal resolution, modeling fidelity, or the number of decision stages increase, decomposition is generally inevitable (Table~\ref{tab:TEP}). Nevertheless, no method is uniformly best because their performance highly depends on the uncertainty formulation, operational details, integrality, and separability.}

\vspace{0.15cm}
\noindent \textbf{Decomposition Algorithms.} {BD and its variants are a natural choice when fixing a set of complicating variables decomposes the model into several independent subproblems \citep{rahmaniani_benders_2017}. Multicut variants can exploit scenario separability and parallel computing, whereas stabilization can reduce oscillation between investment decisions from one iteration to another \citep{Zhang2024_COR_GTEP}. If operational uncertainty unfolds over multiple stages, SDDP avoids explicit enumeration of the complete scenario tree by constructing approximations of future-cost functions, making it suitable for long-horizon problems with convex stage problems, especially in hydrothermal systems \citep{Fullner2025_survey_SDDP, Hole2025_EJOR_GTEP}.  } 

{CCG is generally preferable for two- or three-level RO and DRO formulations in which an adversarial subproblem identifies a worst-case realization and the resulting scenario and recourse decisions are added to the master problem \citep{Gamboa2021_wDRO}. Its effectiveness, therefore, significantly depends on whether the adversarial subproblem can be reformulated and solved efficiently. DW or column generation is more suitable for block-angular models whose many independent subproblems are linked by relatively few coupling constraints \citep{El2023_DW}. PHA can be suitable for scenario-separable SPs and parallel implementation, but for MILP models it is used as an approximate solution. ADMM is useful when separability follows organizational boundaries and parties wish to exchange partial information, as in decentralized multi-vector planning.  
Overall, the choice of solution techniques is driven more by mathematical structure than by the application. }

Recent studies demonstrate the scalability of such methods.  \citet{Zuluaga2024_EPSR_GTEP} apply a variant of PHA for a GTEP problem formulated as a two-stage SP, and solve instances with 8000 nodes and 360 representative days (each day is treated as a scenario). \citet{Zhang2024_COR_GTEP} consider a similar problem and solve instances with up to 4.5 billion constraints and 1 billion variables using a tailored BD. \citet{LeeSun2025_GTEP} solve instances with 52 scenarios (each scenario is a week of hourly operations) and 1493 nodes using a tailored level-bundle method that can handle billions of potential line contingencies. These advances suggest that decomposition techniques will remain essential for future CEMs, especially as model include multiple energy vectors, stakeholders, and market structures \citep{Kaya2025_EJOR_survey}.

\vspace{0.15cm}
\noindent \textbf{Heuristic and Approximation Frameworks.} Heuristics can complement decomposition methods by exploiting problem structure to obtain approximate solutions more efficiently. \citet{Neumann2019_heuristic} propose a heuristic framework based on sequential LP and relaxation of integer variables for a deterministic TEP. \citet{Khorramfar2025_EJOR} develop an approximation framework for a stochastic GTEP based on the observation that relaxing certain constraints significantly reduces the computational burden. 
Depending on the model structure and uncertainty representation, various heuristics can be developed for solution efficiency.

\vspace{0.15cm}
\noindent \textbf{GPU-Acceleration.} Advances in graphics processing units (GPUs) provide another way to improve scalability. It offers a powerful platform for parallelizing linear algebraic operations such as matrix factorization and constraint evaluation. In power systems, GPU-accelerated optimization has already shown promising potential in problems such as alternating current optimal power flow (ACOPF) \citep{Shin2024_GPU} and contingency-constrained DCOPF \citep{Degleris2024gpu} with speedups in the orders of magnitude. 

\vspace{0.15cm}
\noindent \textbf{Quantum Computing.} Some studies attempt to solve MILPs with quantum computing, a technology whose large-scale, fault-tolerant implementation remains a longer-term prospect. A common theme is to reformulate the problem as a quadratic unconstrained binary optimization (QUBO) and apply quantum annealing \citep{morstyn2022annealing, Morstyn2024_survey_QC_power}. 
Hybrid approaches are also popular, where a quantum computer is only used for certain subroutines and the remaining algorithm is classical \citep{Glover2022_QC, Quinton2025_QC, Singha2025_QC_RPS}. However, current theoretical speedups are limited (often quadratic at best) and practical implementations are constrained by the need for large-scale and fault-tolerant quantum hardware \citep{abbas2024challenges, babbush2025grand}.
As a result, quantum computing for solving near-term energy planning is a highly speculative direction for CEMs, though future breakthroughs in algorithms or hardware could change this outlook.


\subsection{Validation and Decision Impact}\label{ssec:validation-decision-impact}
Despite the substantial methodological progress, a fundamental question looms: when and for whom do modeling enhancements actually improve decision-making outcomes? Much of the literature implicitly assumes that increasing modeling fidelity necessarily leads to better planning insights. However, systematic evidence supporting this assumption is scarce. Therefore, structural frameworks are needed to answer the following dimensions.  
\begin{itemize}[nosep]
    \item \textbf{Decision value:} Does a particular modeling choice alter a specific decision for a given class of decision-maker (see Section~\ref{sec:decision-context})?
    \item \textbf{Fidelity threshold:} At what spatio-temporal or operational resolution do key decisions stabilize, such that additional modeling detail leads to marginal change in the outcome?
    \item \textbf{Context boundaries:} Under what planning context does a modeling enhancement significantly affect results, and when does it primarily increase computational burden without improving decision quality?    
\end{itemize}
\vspace{0.1cm}
Several methodological practices can help establish such validation practices. One approach is sensitivity analysis under different spatio-temporal resolutions and operational fidelity while tracking changes in decision outcomes. Another approach is retrospective validation where model outcomes are compared against historical planning decisions. The third approach involves decision-maker experiments in which planners, regulators, and investors are presented with outputs generated by alternative outcomes and their responses are systematically evaluated. {Modeling to generate alternatives (MGA) offers a complementary framework for such experiments. Rather than reporting only one cost-optimal portfolio, MGA introduces an acceptable deviation from the optimal objective and searches for diverse infrastructure outcomes within the resulting near-optimal region. This can offer decision flexibility that is hidden by a single optimum, particularly when relevant considerations (e.g., social acceptance or siting preferences) are difficult to parameterize. Combining MGA with scenario analysis, out-of-sample testing, or uncertainty-aware optimization can therefore identify multiple planning strategies that achieve a desired level of resilience rather than one hedging strategy \citep{Lombardi2025_joule_MGA}}.
{Recent advances further improve the coverage and practical use of these alternatives, allowing fast and interactive generation of customized options \citep{Lau2025_MGA, Turan2026_oracle_MGA}. }

\section{The Role of Machine Learning}\label{sec:ml_role}
{Section~\ref{sec:gaps} raised key challenges and research gaps for energy infrastructure planning under uncertainty: modeling fidelity, uncertainty characterization, solution approaches, and validation of decision impact. While the field has progressed steadily towards solving larger and more realistic planning models alongside developments in traditional optimization methods, recent advances in artificial intelligence and machine learning (ML) present new opportunities for closing the gap towards more robust, high-resolution, and high-fidelity planning models. In this section, we highlight three points where ML is beginning to enter the planning pipeline: forecasting and projection of uncertain parameters (Section~\ref{sec:forecasting}), generative and adversarial scenario generation for out-of-sample robustness (Section~\ref{sec:adverarial}), and learned surrogates for high-fidelity operational modeling (Section~\ref{sec:surrogates}).}

\subsection{Forecasting, Projection, and Distribution Shift}
\label{sec:forecasting}
Forecasting predicts future parameter values from historical data, while projection estimates them under specified assumptions or scenarios. Both are fundamental to energy infrastructure planning because investment decisions depend on long-term trajectories of demand, supply, and technology costs. Traditional econometric forecasting methods have long been used to estimate such trends  \citep{Suganthi2012_sruvey_forecast}. These models are interpretable and suitable for macro-scale projections such as load growth, fuel costs, and technology adoption. However, their simplified structures may struggle to capture the nonlinear dynamics and volatility in modern energy systems with a high VRE share. 

As energy systems generate increasingly large and multi-dimensional datasets, ML methods have shown strong performance in short-term electricity demand forecasting tasks. Models such as gradient boosting \citep{Zhang2023_XGBoost} and recurrent neural network \citep{Bedi2019, Qureshi2024} can capture nonlinear interactions between load, weather conditions, and macroeconomic variables, thereby improving forecasting intraday and weekly horizons. Probabilistic ML approaches, including Bayesian neural networks and Gaussian process models, can also provide predictive distributions that help generate operational scenarios for short-term uncertainty \citep{Pineda2016_EJOR_GTEP}.

At longer horizons, however, ML models may suffer from distribution shifts, as they typically rely on historical data patterns. Structural changes, such as climate-driven shifts in wind resources, can make historical data less representative of future conditions \citep{hdidouan2017}, thus leading to suboptimal infrastructure decisions. To address these challenges, long-term projections may adopt hybrid approaches that combine ML and structural models that ensure forecasts remain consistent with engineering constraints, economic trends, and technology learning dynamics. Meanwhile, scenario analysis and expert judgment can capture policy and technological transitions that historical data cannot fully represent.

Several research challenges are posed. One concerns integrating uncertainty quantification from heterogeneous forecasting models (climate projections and short-term ML forecasts) into coherent scenario representations for CEMs. Another is developing robust validation frameworks, including backtesting under extreme conditions. Progress in energy infrastructure planning applications will therefore depend on combining ML methods with economic and physical insights to produce reliable projections with well-characterized uncertainty estimates.

\subsection{Adversarial Scenario Generation for Out-of-Sample Robustness}
\label{sec:adverarial}
Stochastic and robust CEMs aim to identify planning decisions that remain resilient under uncertain future conditions, including extreme events such as renewable droughts, fuel supply disruptions, and weather-induced grid failure. However, future extremes are expected to differ from those observed historically, both in frequency and severity, as climate change alters the statistical properties of weather and climate systems \citep{hansen2012perception}. Therefore, historical data may no longer represent future risks, leading to outcomes that appears robust but perform poorly under plausible future extremes. This highlights the need for scenario generation methods that capture both rare events and distributional shifts. 

In RO, common uncertainty set representations (e.g, polyhedral) are simple to construct but can result in overly conservative decisions. Recent studies has sought to learn tighter sets from data using methods such as clustering and dimensionality reduction \citep{hong2021learning}, and unsupervised learning techniques \citep{Goerigk2023_COR_RO_NN, chenreddy2022data}. More recently, generative ML approaches have been explored to create adversarial scenarios \citep{Brenner2025_GTEP}. Other work proposes adversarial sample generation through min-max optimization or distributionally robust training of generative models \citep{cheng2025worstcasegenerationminimaxoptimization, dai2025assured}. 


Despite their promise, generative methods face important challenges. Generated scenarios may be statistically plausible but physically inconsistent or fail to capture truly novel extremes if trained only on historical data. Moreover, synthetic scenarios cannot be validated against future observations, making their credibility difficult to assess. To mitigate these limitations, scenario generation should incorporate physical constraints, leverage hybrid approaches with climate or system models, and involve expert review. 

\subsection{Learned Surrogates for High-fidelity Operational Modeling}
\label{sec:surrogates}
Embedding high-fidelity operational models (e.g., security constraint UC) can greatly improve CEM outcomes at the expense of computational burden (see Sec.~\ref{ssec:fidelity}). Even with decomposition algorithms, solving the resulting problem can become prohibitively expensive. ML-surrogate models have recently emerged as a promising alternative to approximate operational behavior at much reduced computational cost. Although not widely applied in CEMs, related work in SP and RO shows the potential of these surrogate models, where they map from first-stage decisions to second-stage operational costs or actions and embed these approximations within optimization algorithms \citep{patel2022neur2sp, dumouchelle2023neur2ro, bertsimas2024machine}. 
{Extending this approach directly to large-scale GTEPs formulated as two-stage SP, \citet{Kwon2027} propose a framework that adaptively selects the scenario count and operational horizon, trains surrogates of the expected scenarios (operational problems), and embeds them in the planning model to obtain high-quality investment plans.  }

A related body of research focuses on single-stage power system optimization, where ML models learn direct mappings from system parameters---such as nodal load profiles---to optimal operational decisions. {Multiple studies propose learned models for AC optimal power flow \citep{donti2021dc3, huang2021deepopf, pan2022deepopf, nguyen2025fsnet, qiu2024dual}. Other studies introduce graph neural network architectures that accommodate varying network topologies \citep{wu2025universal, piloto2024canos, zhou2022deepopf}}; a more comprehensive listing of papers on ML for single-stage power grid optimization is available at \cite{MLOPFwiki2026}. Collectively, this literature suggests that surrogate modeling could enable planning models that retain high-fidelity operational features,
while mitigating the associated computational complexity.




\section{Summary and Discussion}\label{sec:discussion}
The structural pattern of the literature is presented in Fig.~\ref{fig:co-occurrence} which reports the conditional co-occurrence between modeling paradigms, uncertainty characterization, modeling features, and solution approaches. The figure shows how certain modeling choices tend to bundle together, and which combinations remained unexplored. Accordingly, Fig.~\ref{fig:co-occurrence} synthesizes  Sections~\ref{sec:gaps} and \ref{sec:ml_role} to provide a \textit{map} to explore the current literature and identify the persistent gaps quickly.

\subsection{Dominant Modeling Patterns}
\textbf{Mixed-integer stochastic programs are predominant.} Hourly resolution shows strong co-occurrence with decomposition methods (CCG, Benders) and MILP formulations. This is because modeling VRE's operations needs a better chronological representation, but it quickly necessitates decomposition methods. As discussed in Section~\ref{ssec:sol_approach}, computational considerations increasingly shape modeling choices by making a tradeoff between high spatio-temporal resolution and detailed operational fidelity. 

\noindent\textbf{Reverse relationship between spatial and temporal resolution.} Papers with hourly resolution tend to have less spatial granularity than those operating on coarser timescale. Moreover, representative days appear in the majority of the hourly models compared to coarser resolution models. 
These inverse relationships demonstrate the fidelity-tractability frontier, breaking which requires algorithmic innovations (Section~\ref{ssec:sol_approach}) or assessing when both dimensions matter (Section~\ref{ssec:validation-decision-impact}).


\subsection{Underexplored Intersections}
\textbf{Multi-vector planning remains sparse.} There is relatively weak integration of gas and hydrogen systems (alternative vectors). 
Given the critical interconnection between energy vectors (Section~\ref{ssec:fidelity}), the limited co-occurrence suggests a critical research frontier to develop tractable models that capture these interdependencies under a cross-sector uncertainty.

\noindent\textbf{Transmission expansion models overlook policy and reliability features.} The heatmap exposes an imbalance in how policy and reliability features are embedded across problem classes. While models with generation expansion (GEP, GTEP) present stronger co-occurrence with environmental constraints (e.g., RPS, carbon tax, decarbonization constraints) and risk measures, models with transmission expansion (TEP and GTEP) are less consistently bundled with these features. It is important to highlight this asymmetry because transmission is frequently the binding constraint for renewable integration; consequently, if TEP studies systematically overlook decarbonization instruments and risk metrics, they can produce insights that are difficult to translate into practice. Relatedly, risk measures and reliability constraints co-occur more frequently with DRO formulations than expected cost SPs. However, their connection to broader modeling features such as storage and UC appears relatively weak. In other words, rather than an indicator metric, risk aversion needs to be co-designed with operational and flexibility levers to deliver resilience under extreme events. 

\textbf{Opportunities for long-term uncertainty modeling.} The figure highlights that uncertainty is mainly considered in a few dominant parameters, including demand and wind capacity factor, which frequently co-occur with SP and RO formulations. Solar uncertainty appears less dominant, and power price and technology cost uncertainty remain underexplored despite their importance for long-horizon investment decisions. Water inflow uncertainty is rare, suggesting that hydro-climatic uncertainty is still a niche thread despite its strategic importance for adequacy and seasonal risks \citep{Xie2025}. A key opportunity, therefore, is to introduce new paradigms such as DRO while broadening the uncertainty portfolio. 


\textbf{Closing the gap with ML.} Finally, Section~\ref{sec:ml_role} argues that machine learning acts as an enabling layer for several of the weak links highlighted by the heatmap and can reshape the field. So far, ML is rarely integrated as a structural component of planning models and is mainly used for forecasting inputs or dimensionality reduction. Nevertheless, the gaps depicted in Fig.~\ref{fig:co-occurrence} ---such as limited uncertainty portfolio, tail risk modeling, weak multi-vector modeling, and computational bottlenecks--- are areas where ML can be transformative. 

\section{Conclusion}\label{sec:conclusion}
This paper systematically surveyed recent developments in energy systems planning under uncertainty and diagnosed methodological and structural research gaps in the field. Methodologically, it includes a unified review of uncertainty characterization paradigms applied to different capacity expansion classes, a comparative assessment of modeling fidelity across spatio-temporal and operational dimensions, and an analysis of solution techniques. Structurally, it includes a co-occurrence mapping of the literature that identifies dominant modeling features and underexplored intersections across uncertainty types, operational features, and computational architecture. By combining thematic review with gap analysis, this paper aims to provide both a consolidated reference for researchers and a roadmap for advancing scalable, risk-aware, and multi-sector decision-making in energy infrastructure planning. 

\setlength{\bibsep}{3pt plus 0.3ex}
{\footnotesize
\bibliographystyle{apalike}
\bibliography{Bibliography}}

\end{document}